\documentclass{mscs}

\usepackage{mycls}

\title{A general proof certification framework\\for modal logic}

\author{
	Tomer Libal$^1$, Marco Volpe$^2$
\\
		$^1$ The American University of Paris, France, $^2$ fortiss GmbH, Munich, Germany
}

\begin{document}

\maketitle

\begin{abstract}
  One of the main issues in proof certification is that different theorem provers, even when designed for the same logic,
  tend to use different proof formalisms and produce outputs in different formats.
  The project ProofCert promotes the usage of a common specification language and of a small and trusted kernel
  in order to check proofs coming from different sources and for different logics.
  By relying on that idea and by using a classical focused sequent calculus as a kernel,
  we propose here a general framework for checking modal proofs. We present the implementation of the framework in
  a Prolog-like language and show how it is possible to specialize it in a simple and modular way in order to cover
  different proof formalisms, such as labeled systems, tableaux, sequent calculi and nested sequent calculi.
  We illustrate the method for the logic K by providing several examples and discuss how to further extend the approach.
\end{abstract}

\pagestyle{empty}

\section{Introduction}
\label{sect:introduction}

The main difficulty in having general and comprehensive approaches to proof checking and proof certification derives from the fact that proof evidences, even for a single, specific logic, are produced by using several different proof formalisms and proof calculi.
This is the case both for human-generated proofs and for proofs provided by automated theorem provers, which moreover tend to produce outputs in different formats.
Addressing such an issue is one of the goals of the project \pcert\ \cite{erc}. By using well-established concepts of proof theory,
\pcert\ proposes \emph{foundational proof certificates} (\fpc) as a framework
to specify proof evidence formats. Describing a format in terms of an \fpc
allows software to check proofs in this format over a small kernel.

\textsf{Checkers}~\cite{chihaniLR15} is a generic proof certifier based on the \pcert\ ideas.
It allows for the certification of arbitrary proof evidences using various trusted kernels, like the focused classical sequent calculus \LKF~\cite{LiaMil09}.
Such kernels are enriched with additional predicates, which allow more control on the construction of a proof. Dedicated \fpc specifications can be defined, over these predicates, in order to interpret the information coming from a specific proof evidence format, so that the kernel is forced to produce a proof that mirrors, and thus certifies in case of success, the original one.

Different kernels, though, offer different levels of confidence in the correctness of the proof. An important
quality of a kernel is that it is as small as possible. The idea behind it - called the ``de Bruijn Criterion'' \cite{automath70} -
is that small and simple kernels offer higher trust. The kernel employed in this paper, based on $\LKF$, consists in 93 lines of $\lambda$Prolog code and is the same used in other ProofCert publications (e.g.,~\cite{chihaniLR15,LibalV16,ChiMilRen16}) \footnote{When calculating the size of the program, we placed each atomic predicate on a new line.}.

The problem of the great variety of different proof formalisms and proof systems to be considered, when dealing with proof checking, is especially apparent in the case of modal logics, whose proof theory is notoriously non-trivial. In fact, in the last decades several proposals have been provided (a general account is, e.g., in~\cite{Fit07}).
Such proposals range over a set of different proof formalisms
(e.g.,~sequent, nested sequent, labeled	sequent, hypersequent
calculi, semantic tableaux), each of them presenting its own features and drawbacks.
Several results concerning correspondences and connections between the
different formalisms are also known~\cite{Fit12,GorRam12,Lel15}.

In~\cite{MarMilVol16}, a general framework for emulating and comparing existing modal proof systems has been presented.
Such a framework is based on the setting of \emph{labeled deduction
  systems}~\cite{Gab96}, which consists in enriching the syntax of modal logic with elements coming from the semantics, i.e., with elements referring explicitly to the worlds of a Kripke model and to the accessibility relation between such worlds.
In particular, the framework is designed as a focused version of Negri's system G3K~\cite{Neg05}, further enriched with a few parametric devices.
Playing with such parameters produces concrete instantiations of the framework, which, by exploiting the expressiveness of the labeled
		approach and the control mechanisms of focusing, can be used to emulate the behavior of a range of existing formalisms and proof systems for modal logic with high precision.

In this paper, we rely on the close relationship between labeled sequent systems and $\LKF$~\cite{MilVol15} in order to propose an implementation of such a framework that uses $\LKF$ as a kernel, and is developed as a module of the more general \textsf{Checkers} implementation project.
This work also capitalizes on (and, in a sense, generalizes) the one in~\cite{LibalV16}, which was limited to the case of prefixed tableaux.
The implementation is extremely modular and based on the use of layers that mirror quite closely the instantiations of the framework presented in~\cite{MarMilVol16}.
Concretely, we are able to certify, via this implementation, proofs given in the formalisms of labeled sequents, prefixed tableaux, ordinary sequent systems and nested sequents.
We cover for the moment only the modal logic $K$, but the modularity of the approach should allow for an easy extension to other modal logics, in particular those whose Kripke frames are defined by geometric axioms, according to the treatment described in~\cite{MarMilVol16}.
Extension to other formalisms seems also possible; we discuss this in more detail in the conclusion.

An approach related to ours is in~\cite{benz2015}, where the authors present a technique to generate and certify modal proofs using the Coq proof assistant.
The aim of their work is to allow interactive theorem proving over higher-order modal logics. To this end, they encode the semantics
of higher-order modal logics into the system used by Coq -- the Calculus of Inductive Constructions. Their work and ours are similar in that they both
certify modal logic proofs by using trusted kernels, but they also differ in several ways.
Their system targets higher-order modal logics and is also directed towards interactive theorem proving, while ours is for the moment restricted to the task of certification for propositional modal logic. On the other hand, while their encoding focuses on one specific proof format and calculus, we aim, via our framework, at supporting different formalisms and proof systems.


We proceed as follows.
In Sec.~\ref{sec:background}, we present some background on ProofCert, modal logic and proof systems for modal logic.
In Sec.~\ref{sec:framework-aiml}, we recall the general framework of~\cite{MarMilVol16}.
In Sec.~\ref{sec:cert}, we describe its implementation, by presenting the \fpc specifications of the different layers and by providing a few examples.
In Sec.~\ref{sec:conclusion}, we
discuss possible directions for future work, compare with some related approaches, and conclude.

\section{Background}
\label{sec:background}

\subsection{Proof systems for modal logic}

In this section, we review several proof systems that are among the most popular calculi~\cite{Fit07} for automated theorem proving in modal logic as well
as for manual proof generation. Before that, we recall a few key notions about modal logic and its relation with first-order classical logic.

We remark that throughout this paper, we will work with formulas in \emph{negation normal form}, i.e., such that only atoms may possibly occur negated in them. Notice that this is not a restriction, as it is always possible to convert a propositional formula into an equivalent formula in negation normal form (both in classical and in modal logic).

\subsubsection{Modal logic}

The language of \emph{(propositional) modal formulas} consists of a functionally complete set of classical propositional connectives, a \emph{modal operator} $\square$ (here we will also use explicitly its dual $\lozenge$) and a denumerable set $\cal{P}$ of \emph{propositional symbols}. The grammar is specified as follows:
\small
  \begin{displaymath}
  A ::= \, P \, \mid \, \neg P \, \mid \, A \vee A \, \mid \, A \wedge A \, \mid \, \square A \, \mid \, \lozenge A \,
  \end{displaymath}
  \normalsize
where $P \in \cal{P}$. We say that a formula is a \emph{$\square$-formula} (\emph{$\lozenge$-formula}) if its main connective is $\square$ ($\lozenge$).
  The semantics of the modal logic $\logick$ is usually defined by means of \emph{Kripke frames}, i.e.,~pairs $\mathcal{F}=(\wld,\rel)$ where $\wld$ is a non-empty set of \emph{worlds} and $\rel$ is a binary relation on $\wld$. We say that a world \emph{$w$ is accessible from a world $w'$} iff $(w,w')\in \rel$. A \emph{Kripke model} is a triple $\m=(\wld, \rel, \val)$ where $(\wld,\rel)$ is a Kripke frame and $\val: \wld \rightarrow 2^\prop$ is a function that assigns to each world in $\wld$ a (possibly empty) set of propositional symbols.

In the basic modal logic $K$, we define the \emph{truth} of a modal formula at a world $w$ in a Kripke model $\m=(\wld,\rel,\val)$ as the
smallest relation $\models$ satisfying:
\small
	\begin{eqnarray*}
		\m, w \models P
		& \quad \text{iff} \quad & P \in \val(w) \\
		\m, w \models \neg P
		& \quad \text{iff} \quad & P \not\in \val(w) \\
		\m, w \models A \vee B
		& \quad \text{iff} \quad & \m, w \models A
		\text{ or } \m, w \models B \\
		\m, w \models A \wedge B
		& \quad \text{iff} \quad & \m, w \models A
		\text{ and } \m, w \models B \\
		\m, w \models \square A
		& \quad \text{iff} \quad & \m, w' \models A
		\text{ for all } w' \text{ s.t. } w \rel w'\\
		\m, w \models \lozenge A
		& \quad \text{iff} \quad & \text{ there exists } w' \text{ s.t. } w \rel w'	\text{ and } \m, w' \models A.
	\end{eqnarray*}
\normalsize
 By extension, we write
$\m \models A$ when $\m,w \models A \mbox{ for all } w \in \wld$ and
 we write $\models A$ when $\m \models A$ for every Kripke model $\m$.

\subsubsection{The standard translation from modal logic into classical logic}
\label{sec:std-translation}

The following \emph{standard translation} (see, e.g.,~\cite{BlaVBe07}) provides a bridge between propositional (classical) modal logic and first-order classical logic:
\begin{center}
\begin{tabular}{ccc@{\qquad\qquad}ccc}\small
	$\str{P}{x}$ &\small = & \small$P(x)$ &
\small	$\str{A \wedge B}{x}$ & \small= &\small $\str{A}{x} \wedge \str{B}{x}$\\
\small	$\str{\neg P}{x}$ &\small = & \small$\neg P(x)$ &
	\small$\str{\square{A}}{x}$ & \small= &\small $\forall y (\relfo(x,y) \impl \str{A}{y})$\\
\small	$\str{A \vee B}{x}$ & \small= &\small $\str{A}{x} \vee \str{B}{x}$ &
	\small$\str{\lozenge A}{x}$ &\small = &\small $\exists y (\relfo(x,y) \wedge \str{A}{y})$
\end{tabular}
\end{center}
where $x$ is a free variable denoting the world in which the formula is being evaluated. The first-order language into which modal formulas are translated is usually referred to as \emph{first-order correspondence language}~\cite{BlaVBe07} and consists of a binary predicate symbol $\relfo$ and a unary predicate symbol $P$ for each $P\in \prop$. When a modal operator is translated, a new fresh variable is introduced.\footnote{In fact, it is possible to show that every modal formula can be translated into a formula in the fragment of first-order logic which uses only two variables~\cite{BlaVBe07}. By the decidability of such a fragment, an easy proof of the decidability of the modal logic $K$ follows.}
It is easy to show that for any modal formula $A$, any model $\m$ and any world $w$, we have that $\m,w \models A$ if and only if $\m \models \str{A}{x}[x\leftarrow w]$.

\subsubsection{Labeled sequent systems}

	 Several different deductive formalisms have been used for modal proof theory and theorem proving. One of the most interesting approaches has been presented in~\cite{Gab96} with the name of labeled deduction.
	 The basic idea behind labeled proof systems for modal logic is to internalize elements of the corresponding Kripke semantics (namely, the worlds of a Kripke model and the accessibility relation between such worlds) into the syntax.
	 A concrete example of such a system is the sequent calculus $´ G3K$ presented in \cite{Neg05} (we present it here in a single-sided formulation and refer to it as $\labk$).
	 \emph{$\labk$ formulas} are either \emph{labeled formulas} of the form $x:A$ or \emph{relational atoms} of the form $x \rel y$, where $x, y$ range over a set of variables and $A$ is a modal formula. In the following, we will use $\varphi, \psi$ to denote $\labk$ formulas.
	\emph{$\labk$ sequents} have the form $\ar \vdash \Delta$, where $\Delta$ is a multiset containing labeled formulas and $\ar$ is a
  set of relational atoms.
	Being $\labk$ a labeled system, we say that $A$ is provable in $\labk$ if there is a proof of $\vdash x:A$ for any variable $x$.
	In Fig.~\ref{fig:labk}, we present the rules of $\labk$, which is proved to be sound and complete for the basic modal logic $\logick$~\cite{Neg05}.
		\begin{figure}[t]\small
		{\sc Classical rules}
		\[
		\infer[\init_{LS}]{{\ar\vdash \Delta, x:\neg P, x:P}}{}
		\qquad
		\infer[\wedge_{LS}]{\ar\vdash \Delta, x:A \wedge B}{\ar\vdash \Delta, x:A & \ar \vdash \Delta, x:B}
		\quad
		\infer[\vee_{LS}]{\ar\vdash \Delta, x:A \vee B}{\ar\vdash \Delta, x:A, x:B}
		\]
		{\sc Modal rules}
		\[
		\infer[\Box_{LS}]{\ar\vdash \Delta, x:\square A}{\ar\cup\{x \rel y\}\vdash \Delta, y:A}
		\quad
		\infer[\Diamond_{LS}]{\ar\cup\{x\rel y\}\vdash \Delta, x:\lozenge A}{\ar\cup\{x \rel y\}\vdash \Delta, x:\lozenge A, y:A}
		\]
		In $\Box_{LS}$, $y$ does not occur in the conclusion.
		\caption{$\labk$: a labeled sequent system for the modal logic $\logick$.}
		\label{fig:labk}
		\end{figure}

\subsubsection{Prefixed tableau systems}
	\label{sec:fitting-tableaux}
Prefixed tableaux ($\pt$) can also be seen as a particular kind of labeled deductive system. They were introduced in~\cite{fitting1972tableau}, although the formulation that we use here is closer to the one in~\cite{Fit07}. Differently from $\labk$, $\pt$ are refutation proof systems, i.e., in order to prove a formula, we negate it and derive from it a contradiction.
A \emph{prefix} is a finite sequence of positive integers (written by using dots as delimiters). Intuitively, prefixes denote possible worlds and they are such that if $\sigma$ is a prefix, then $\sigma.1$ and $\sigma.2$ denote two worlds accessible from $\sigma$. A \emph{prefixed formula} is $\sigma:A$, where $\sigma$ is a prefix and $A$ is a modal formula in negation normal form.
A prefixed tableau proof of $A$ starts with a root node containing $1:A$, informally asserting that $A$ is false in the world named by the prefix $1$. It continues by using the branch extension rules given in Fig.~\ref{fig:fit-rules}. We say that a branch of a tableau is a \emph{closed branch} if it contains $\sigma:P$ and $\sigma:\neg P$ for some $\sigma$ and some $P$. The goal is to produce a \emph{closed tableau}, i.e.,~a tableau such that all its branches are closed.
		\begin{figure}[tb]\small
		{\sc Classical rules}
		\[
		\infer[\wedge_{PT}]{\sigma: A,\,\sigma:B}{\sigma: A \wedge B}
		\qquad
		\infer[\vee_{PT}]{\sigma: A \quad \mid \quad \sigma:B}{\sigma: A \vee B}
		\]
		{\sc Modal rules}
		\[
		\infer[\square_{PT}]{\sigma.n:A}{\sigma:\square A}
		\qquad
		\infer[\lozenge_{PT}]{\sigma.n:A}{\sigma:\lozenge A}
		\]
		In $\square_{PT}$, $\sigma.n$ is used. In $\lozenge_{PT}$, $\sigma.n$ is new.
		\caption{$\pt$: a prefixed tableau system for the modal logic $\logick$.}
		\label{fig:fit-rules}
		\end{figure}
Classical rules in Fig.~\ref{fig:fit-rules} are the prefixed version of the standard ones. For what concerns the modal rules, the $\lozenge$ rule applied to a formula $\sigma:A$ intuitively allows for generating a new world, accessible from $\sigma$, where $A$ holds, while the $\square$ rule applied to a formula $\square:A$
allows for moving the formula $A$ to an already existing world accessible from $\sigma$.

\subsubsection{Ordinary sequent systems}
\label{sec:bg-ordinary}

	Several ``ordinary'' sequent systems have been proposed in the literature for different modal logics (a general account is, e.g.,~in \cite{Ind10,Pog11}).
	In our treatment, we will use the formalization $\os$ presented in Fig.~\ref{fig:os}, which is adapted mainly from the presentations  in~\cite{Fit07,SteSto04}.
	The base classical system (consisting of \emph{identity}, \emph{structural} and \emph{classical connective} rules) is extended by a modal rule that, works on one $\square$-formula and several $\lozenge$-formulas, bottom-up.


%
	\begin{figure}
		{\sc Classical rules}\small
		\[
				\infer[\initos]{\vdash{\Gamma, P, \neg P}}{}
				\qquad \qquad
		\infer[\wedge_{OS}]{\vdash \Gamma, A \wedge B}{\vdash \Gamma, A & \vdash \Gamma, B}
		\qquad \qquad
		\infer[\vee_{OS}]{\vdash \Gamma, A \vee B}{\vdash \Gamma, A, B}
		\]
		{\sc Modal rules}
		\[
		\infer[\boxos]{\vdash \lozenge \Gamma, \square A, \Delta}{\vdash \Gamma, A}
		\]
		\caption{$\os$: an ordinary sequent system for the modal logic $K$.}
		\label{fig:os}
	\end{figure}

\subsubsection{Nested sequent systems}
\label{sec:nss}

	Nested sequents (first introduced by Kashima \cite{Kas94}, and then independently rediscovered by Poggiolesi \cite{Pog11}, as \emph{tree-hypersequents}, and by Br\"unnler \cite{Bru09}) are an extension of ordinary sequents to a structure of tree, where each $[\,]$-node represents the scope of a modal $\square$.
	We write a nested sequent as a multiset of formulas and \emph{boxed sequents}, according to the following grammar, where $A$ can be any modal formula in negative normal form:
	$
	\mathcal N ::= \emptyset \mid A, \mathcal N \mid [\mathcal N], \mathcal N
	$

	In a nested sequent calculus, a rule can be applied at any depth in this tree structure, that is, inside a certain nested sequent context.
	A \emph{context} written as $\Gcon{\ }\cdots\{\ \}$ is a nested sequent with a number of holes occurring in place of formulas (and never inside a formula).
	Given a context $\Gcon{\ }\cdots\{\ \}$ with $n$ holes, and $n$ nested sequents $\mathcal M_1, \ldots, \mathcal M_n$, we write $\Gcon{\mathcal M_1}\cdots\{\mathcal M_n\}$ to denote the nested sequent where the $i$-th hole in the context has been replaced by $\mathcal M_i$, with the understanding that if $\mathcal M_i = \emptyset$ then the hole is simply removed.
	We are going to consider the nested sequent system (in Fig.~\ref{fig:ns}) introduced by Br\"unnler in~\cite{Bru09}, that we call here $\nss$.

	\begin{figure}
		{\sc Classical rules}\small
		\[
		\infer[\init_{NS}]{\Gcon{P, \neg P}}{}
		\qquad \qquad
		\infer[\wedge_{NS}]{\Gcon{A \wedge B}}{\Gcon{A} & \Gcon{B}}
		\qquad 
		\infer[\vee_{NS}]{\Gcon{A \vee B}}{\Gcon{A, B}}
		\smallskip
		\]
		{\sc Modal rules}
		\[
		\infer[\square_{NS}]{\Gcon{\square A}}{\Gcon{[A]}}
		\qquad
		\infer[\lozenge_{NS}]{\Gcon{\lozenge A, [\mathcal M]}}{\Gcon{\lozenge A, [A, \mathcal M]}}
		\]
		\caption{$\nss$: a nested sequent system for the modal logic $K$.}
		\label{fig:ns}
	\end{figure}

\subsection{A general proof checker}
\label{sec:checkers}

There is no consensus about what shape a formal proof evidence should take. The notion of structural proofs, which is based
on derivations in some calculus, is of no help as long as the calculus is not fixed. One of the ideas of the \pcert\ project
is to try to amend this problem by defining the notion of a foundational proof certificate (\fpc) as a pair of an arbitrary
proof evidence and an executable specification which denotes its semantics in terms of some well known target calculus,
such as the sequent calculus. These two elements of an \fpc\ are then
given to a universal proof checker which, by the help of the \fpc, is capable of deriving a proof in the target calculus.
Since the proof generated is over a well known and low-level calculus which is easy to implement, one can obtain
a high degree of trust in its correctness. Such an approach seems to be applicable to a large class of proof formalisms.

The proof certifier \textsf{Checkers} is a $\lambda$Prolog \cite{Miller2012} implementation of this idea.
Its main components are the following:

\begin{itemize}
  \item {\bf Kernel.} The kernels are the implementations of several trusted proof calculi. Currently, there
    are kernels over the classical and intuitionistic focused sequent calculus. Sec.~\ref{sec:lkf} is
  devoted to the presentation of \LKF{}, i.e.,~the classical focused sequent calculus that will be used in the paper.
  \item {\bf Proof evidence.} The first component of an \fpc, a proof evidence is a $\lambda$Prolog description
    of a proof output of a theorem prover. Given the high-level declarative form of $\lambda$Prolog, the structure and
    form of the evidence are very similar to the original proof.
    We specify the form of the different proof evidences we handle in Sec.~\ref{sec:cert}.
  \item {\bf \fpc specification.} The basic idea of \textsf{Checkers} is to try and generate a proof of the theorem
    of the evidence in the target kernel. In order to achieve that, the different kernels have additional predicates
    which take into account the information given in the evidence. Since the form of this information is not known
    to the kernel, \textsf{Checkers} uses \fpc specifications in order to interpret it. These
    logical specifications are written in $\lambda$Prolog and interface with the kernel in a sound way in order
    to certify proofs. Writing these specifications is the main task for supporting the different outputs of
    the modal theorem provers we consider in this paper and they are, therefore, explained in detail in Sec.~\ref{sec:cert}.
    We mention here the existence of two different types of specifications. The \emph{clerks}, which simply perform some bookkeeping computations without using any information from the evidence, and the \emph{experts}, which, in addition, also use information from the evidence in order to guide the kernel with regard to choices to make.
\end{itemize}

\subsection{A focused sequent calculus for classical logic}
\label{sec:lkf}

Theorem provers often use efficient but non-trivial proof calculi, possibly employing heuristics or optimization techniques, whose complexity leads to a lower degree of trust. On the other hand, traditional proof calculi, like the
sequent calculus, enjoy a high degree of trust but are quite inefficient for proof search.
In order to use the sequent calculus as the basis of automated deduction, much more structure within proofs needs to be established.
Focused sequent calculi, first introduced by Andreoli \cite{andreoli1992logic} for linear logic,
combine the higher degree of trust of sequent calculi with a more efficient proof search. They take advantage of the fact that
some of the rules are ``invertible'', i.e., can be applied without requiring backtracking, and that some other rules can ``focus''
on the same formula for a batch of deduction steps. In this paper, we will make use of the
classical focused sequent calculus (\LKF) system defined in \cite{LiaMil09}. Fig.~\ref{fig:lkf} presents,
in the black font, the rules of \LKF.

Formulas in \LKF\, which are expressed in negation normal form, can have either positive or negative polarity and are constructed from atomic formulas, whose polarity has to be assigned, and from logical connectives whose polarity is pre-assigned.
The choice of polarization does not affect the provability of a formula, but it can have a big impact on proof search and on the structure of proofs: one can observe, e.g., that in \LKF the rule for $\vee^-$ is invertible while the one for $\vee^+$ is not.
The connectives $\wedge^-,\vee^-$ and $\forall$ are of negative polarity, while $\wedge^+, \vee^+$ and $\exists$ are of positive polarity. A composed formula has the same polarity of its main connective.
In order to polarize literals, we are allowed to fix the polarity
of atomic formulas in any way we see fit. We may ask that all atomic formulas
are positive, that they are all negative, or we can mix polarity assignments. In
any case, if $A$ is a positive atomic formula, then it is a positive formula and $\neg A$
is a negative formula: conversely, if $A$ is a negative atomic formula, then it is a
negative formula and $\neg A$ is a positive formula

Deductions in \LKF\ are done during synchronous or asynchronous phases. A synchronous phase, in which sequents have the form $\sync{\Theta}{B}$, corresponds to the application of synchronous rules to a specific positive formula $B$ under focus (and possibly its immediate positive subformulas). An asynchronous phase, in which sequents have the form $\async{\Theta}{\Gamma}$, consists in the application of invertible rules to negative formulas contained in $\Gamma$ (and possibly their immediate negative subformulas).
Phases can be changed by the application of the ${release}$ rule.
A \emph{bipole} is a pair of a synchronous phase below an
asynchronous phase within \LKF: thus, bipoles are macro inference rules in
which the conclusion and the premises are $\Uparrow$-sequents with no
formulas to the right of the up-arrow.

It is useful sometimes to delay the application of invertible rules (focused rules) on some negative formulas (positive formulas) $A$.
In order to achieve that, we define the following delaying operators
$\delayop^+(A) = \texttt{true} \wedge^+ A$ and $\delayop^-(A) = \texttt{false} \vee^- A$.
Clearly, $A,\delayop^+(A)$ and $\delayop^-(A)$
are all logically equivalent but $\delayop^+(A)$ is always a positive formula and $\delayop^-(A)$ is always a negative one.

In order to integrate the use of \fpc into the calculus, we enrich each rule of \LKF\ with proof evidences and additional predicates,
given in blue font in Fig.~\ref{fig:lkf}. We call the resulted calculus $\aLKF$. $\aLKF$ extends $\LKF$ in the following way.
Each sequent now contains additional information in the form of the proof evidence $\Xi$.
At the same time, each rule is associated with a predicate (for example \renewcommand{\initExpert}[2]{{\hbox{\sl initial$_e$}(#1,#2)}}$\initExpert\Xi l$) which,
according to the proof evidence, might prevent the rule from being called or guide it by supplying such information as
the cut formula to be used.

Note that adding the \fpc\ definitions in Fig.~\ref{fig:lkf} does not harm the soundness of the system but only restricts
the possible rules which can be applied at each step. Therefore, a proof obtained using \aLKF\ is also a proof in \LKF.
Since the additional predicates do not compromise the soundness of \aLKF, we allow their definition to be external to the kernel and in fact
these definitions, which are supplied by the user, are what allow \textsf{Checkers} to check arbitrary proof formats.

\begin{figure}[tb]
\renewcommand{\Async}[3]{\blue{#1}\vdash#2\mathbin{\Uparrow}   #3}
\renewcommand{\Sync }[3]{\blue{#1}\vdash#2\mathbin{\Downarrow} #3}
\renewcommand{\andClerk}[3]{\blue{{\hbox{andNeg$_c$}}(#1,#2,#3)}}
\renewcommand{\falseClerk}[2]{\blue{\hbox{f$_c$}(#1,#2)}}
\renewcommand{\orClerk}[2]{\blue{{\hbox{orNeg$_c$}}(#1,#2)}}
\renewcommand{\allClerk}[2]{\blue{\hbox{all$_c$}(#1,#2)}}
\renewcommand{\storeClerk}[4]{\blue{\hbox{\sl store$_c$}(#1,#2,#3,#4)}}
\renewcommand{\trueExpert }[1]{\blue{{\hbox{true$_e$}}(#1)}}
\renewcommand{\andExpert}[3]{\blue{{\hbox{andPos$_e$}}(#1,#2,#3)}}
\renewcommand{\andExpertLJF}[6]{\blue{{\hbox{andPos$_e$}}(#1,#2,#3,#4,#5,#6)}}
\renewcommand{\orExpert  }[3]{\blue{{\hbox{orPos$_e$}}(#1,#2,#3)}}
\renewcommand{\someExpert}[3]{\blue{\hbox{some$_e$}(#1,#2,#3)}}
\renewcommand{\initExpert}[2]{\blue{\hbox{\sl initial$_e$}(#1,#2)}}
\renewcommand{\cutExpert}[4]{\blue{\hbox{\sl cut$_e$}(#1,#2,#3,#4)}}
\renewcommand{\decideExpert}[3]{\blue{\hbox{\sl decide$_e$}(#1,#2,#3)}}
\renewcommand{\releaseExpert}[2]{\blue{\hbox{\sl release$_e$}(#1,#2)}}

{\sc Asynchronous introduction rules}
\[
\infer{\Async{\Xi}{\Theta}{A\wedgen B,\Gamma}}
      {\Async{\Xi'}{\Theta}{A,\Gamma} \quad
       \Async{\Xi''}{\Theta}{B,\Gamma} \quad
       \andClerk{\Xi}{\Xi'}{\Xi''}}
\]
\[
\infer{\Async{\Xi}{\Theta}{ A\veen B,\Gamma}}
      {\Async{\Xi'}{\Theta}{ A,B,\Gamma}\quad\orClerk{\Xi}{\Xi'}}
\qquad
\infer[\dag]{\Async{\Xi}{\Theta}{ \forall x.B,\Gamma}}
      {\Async{(\Xi' y)}{\Theta}{[y/x]B,\Gamma}\quad\allClerk{\Xi}{\Xi'}}
\]
{\sc Synchronous introdution rules}
\[
\infer{\Sync{\Xi}{\Theta}{B_1\wedgep B_2}}
      {\Sync{\Xi'}{\Theta}{B_1}\quad
       \Sync{\Xi''}{\Theta}{B_2}\quad
       \andExpert{\Xi}{\Xi'}{\Xi''}}
\]
\[
\infer{\Sync{\Xi}{\Theta}{B_1\veep B_2}}{\Sync{\Xi'}{\Theta}{B_i}\qquad
       \orExpert{\Xi}{\Xi'}{i}}
\qquad\qquad
\infer{\Sync{\Xi}{\Theta}{\exists x.B}}{\Sync{\Xi'}{\Theta}{[t/x]B}\quad
                  \someExpert{\Xi}{t}{\Xi'}}
\]
{\sc Identity rules}
\[
\infer[cut]{\Async{\Xi}{\Theta}{\cdot}}
           {\Async{\Xi'}{\Theta}{B}\quad
            \Async{\Xi''}{\Theta}{\neg{B}}
            \quad \cutExpert{\Xi}{\Xi'}{\Xi''}{B}}
\qquad
\infer[init]{\Sync{\Xi}{\Theta}{P_a}}
            {\tupp{l}{\neg P_a}\in\Theta\quad\initExpert{\Xi}{l}}
\]
{\sc Structural rules}
\[
\infer[\kern -1pt release]{\Sync{\Xi}{\Theta}{N}}
               {\Async{\Xi'}{\Theta}{N}\quad\releaseExpert{\Xi}{\Xi'}}
\qquad
\infer[store]{\Async{\Xi}{\Theta}{C,\Gamma}}
             {\Async{\Xi'}{\Theta, \tupp{l}{C}}{\Gamma} \quad
              \storeClerk{\Xi}{C}{l}{\Xi'}}
\]
\[
\infer[\kern -1pt decide]{\Async{\Xi}{\Theta}{\cdot}}
              {\Sync{\Xi'}{\Theta}{P}\quad
               \tupp{l}{P}\in\Theta\quad
               \decideExpert{\Xi}{l}{\Xi'}}
\]
\caption{The proof system \aLKF, augmented version of \LKF.
Here, $P$ is a positive formula; $N$ a negative formula; $P_a$ a positive literal; $C$ a positive formula or
  negative literal; and $\neg B$ is the negation normal form of the negation of B. The proviso marked $\dag$
  requires that $y$ is not free in $\blue{\Xi,}\Theta,\Gamma,B$. \LKF is obtained by ignoring the blue elements in the figure.}
\label{fig:lkf}
\end{figure}

\section{A general focused framework for modal logic}
\label{sec:framework-aiml}

\subsection{A focused labeled calculus for modal logic}
	\label{sec:lmf}


In~\cite{MilVol15}, a focused labeled sequent system ($\lmf$) for the modal logic $K$ has been presented.
Such a calculus can be seen either as a focused version of $\labk$ or as the restriction of $\LKF$ to the first-order correspondence language of Sec.~\ref{sec:std-translation} (where modalities are considered as synthetic connectives).

Fig.~\ref{fig:lmf} presents a multi-focused version (denoted $\lmfm$) of the calculus, i.e., a variant where it is possible to focus on several positive formulas at the same time. Such a variant will also be considered in the rest of the paper.
$\lmf$ can be read from the figure by ignoring the elements in blue font, or, equivalently, by imposing the condition that $\Omega$, $\Omega_1$ and $\Omega_2$ are empty in all rules.

In these systems, sequents have the form $\syncr{\ar}{}{\Theta}{\Gamma}$ (with $\Gamma$ containing exactly one formula in the case of $\lmf$) or $\asyncr{\ar}{}{\Theta}{\Gamma}$, where the \emph{relational set (of the sequent)} $\ar$ is a set of relational atoms and $\Theta$ and $\Gamma$ are multisets of labeled formulas.

\begin{figure}[t]
	{\sc Asynchronous introduction rules}
	\[
	\infer[\wedgenk]{\asyncr{\ar}{}{\Theta}{x:A\wedgen B,\Gamma}}
	{\asyncr{\ar}{}{\Theta}{x:A,\Gamma}\quad \asyncr{\ar}{}{\Theta}{x:B,\Gamma}}
	\quad\quad
	\infer[\kern-2pt\veenk]{\asyncr{\ar}{}{\Theta}{x\colon A\veen\kern-2pt B,\Gamma}}{\asyncr{\ar}{}{\Theta}{x\colon A,x\colon B,\Gamma}}
	\]
	\[
	\infer[\kern-2pt\boxk]{\asyncr{\ar}{}{\Theta}{x:\square B, \Gamma}}{\asyncr{\ar\cup\{x \rel y\}}{}{\Theta}{y\colon B,\Gamma}}
	\]
	{\sc Synchronous introduction rules}
	\[
	\infer[\wedgepk]{\syncr{\ar}{}{\Theta}{x:A\wedgep B\blue{, \Omega_1, \Omega_2}}}
	{\syncr{\ar}{}{\Theta}{x:A\blue{, \Omega_1}}\quad\syncr{\ar}{}{\Theta}{x:B\blue{, \Omega_2}}}
	\]
	\[
	\infer[\kern-3pt\veep, i\in\{1, 2\}]{\syncr{\ar}{}{\Theta}{x:B_1\veep\kern-2pt B_2\blue{, \Omega}}}{\syncr{\ar}{}{\Theta}{x:B_i\blue{, \Omega}}}
	\quad
	\quad
	\infer[\kern-2pt\diamondk]{\syncr{\ar{\cup\{x \rel y\}}}{}{\Theta}{x:\lozenge B\blue{,\Omega}}}{\syncr{\ar\cup\{x \rel y\}}{}{\Theta}{y:B\blue{,\Omega}}}
	\]
	{\sc Identity rules}
	\[
	\infer[\initk]{\syncr{\ar}{}{x:\neg B,\Theta}{x:B}}{}
	\quad
	\]
	{\sc Structural rules}
	\[
	\infer[\storek]{\asyncr{\ar}{}{\Theta}{x:B,\Gamma}}{\asyncr{\ar}{}{\Theta,x:B}{\Gamma}}
	\quad\
	\infer[\releasek]{\syncr{\ar}{}{\Theta}{x:B\blue{, \Omega}}}{\asyncr{\ar}{}{\Theta}{x:B\blue{, \Omega}}}
	\quad\
	\infer[\decidek]{\asyncr{\ar}{}{x:B,\blue{\Omega,}\Theta}{\cdot}}{\syncr{\ar}{}{x:B,\blue{\Omega,}\Theta}{x:B\blue{, \Omega}}}
	\]
	In $\decidek$, $B$ and the formulas in $\Omega$ are positive; in $\releasek$, $B$ and the formulas in $\Omega$ are negative; in $\storek$, $B$ is a positive formula or a negative literal; in $\initk$, $B$ is a positive literal. In $\boxk$, $y$ is different from $x$ and does not occur in
	$\Theta$, $\Gamma$, $\ar$.
	\caption{\lmfm: a multi-focused labeled proof system for the modal logic $\logick$. The single-focused version $\lmf$ is obtained by forcing $\Omega$, $\Omega_1$ and $\Omega_2$ to be empty in all rules.}
	\label{fig:lmf}
\end{figure}

%
%
%

\subsection{A general framework for modal logic}
\label{sec:framework}

In the context of modal logics, labeled proof systems have been shown to be quite expressive and encodings of other approaches into this formalism have also been presented in the literature~\cite{Fit12,GorRam12,Lel15}. It seems therefore quite natural to explore the possibility of reproducing the behavior of modal proof systems based on different formalisms inside $\lmf$, by exploiting at the same time the expressivity of labeling and the control mechanisms provided by focusing.
Such an analysis has been carried out in~\cite{MarMilVol16} and has shown that, by enriching $\lmf$ with a few further technical devices, it is possible to get enough power to drive construction of proofs so as to emulate the proof structure of a wide range of formalisms.




	The general framework $\lmfstar$ is presented in Fig.~\ref{fig:lmfstar}.
	In the rest of this paper, when talking of $\lmfstar$ and its instantiations, a \emph{labeled formula} will have the form $\varphi \equiv x\sigma:A$, where $\sigma$ is either empty or a label $y$. We say that $x$ is the \emph{present} of $\varphi$ and $\sigma$ is the \emph{future} of $\varphi$. Intuitively, the present of a formula has the usual role of labels in labeled systems, i.e.,~it refers to the world where the formula holds. The future, when present, is used to drive (bottom-up) the applications of the $\lozenge$ introduction rule, i.e., it specifies the label to be used as a witness in the rule.
	An \emph{$\lmfstar$ sequent} has the form $\syncr{\ar}{\hp}{\Theta}{\Omega}$ or $\asyncr{\ar}{\hp}{\Theta}{\Omega}$, where $\ar$ is a set of relational atoms, the \emph{present (of the sequent)} $\hp$ is a non-empty multiset of labels, and $\Theta$ and $\Omega$ are multisets of labeled formulas. Intuitively, the present of a sequent specifies which labels are currently ``active'', in the sense that when building a proof (bottom-up) a $\decide$ rule can only put the focus on labeled formulas whose present is contained in the present of the sequent. Like $\lmfm$, $\lmfstar$ is a multi-focused system, as one can notice by observing that we do not necessarily have a single formula on the right of $\Downarrow$.

	We refer the reader to~\cite{MarMilVol16} for a more comprehensive explanation of the devices introduced in $\lmfstar$ with respect to $\lmf$ and $\lmfm$. We just remark that the framework presented here is slightly different from the one proposed in~\cite{MarMilVol16}, since considering only the logic $K$ allows for a few simplifications.

			\begin{figure}
			{\sc Asynchronous introduction rules}
			\[
				\infer[\wedgenf]{\asyncr{\ar}{\hp}{\Theta}{x:A\wedgen B,\Omega}}
				      {\asyncr{\ar}{\hp}{\Theta}{x:A,\Omega}\quad \asyncr{\ar}{\hp}{\Theta}{x:B,\Omega}}
				\qquad
			\infer[\veenf]{\asyncr{\ar}{\hp}{\Theta}{x:A\veen B,\Omega}}{\asyncr{\ar}{\hp}{\Theta}{x:A,x:B,\Omega}}
			\]
			\[
			\infer[\boxf]{\asyncr{\ar}{\hp}{\Theta}{x:\square B, \Omega}}{\asyncr{\ar \cup \{x\rel y\}}{\hp}{\Theta}{y:B,\Omega}}
			\]
	%
			{\sc Synchronous introduction rules}
			\[
			\infer[\wedgepf]{\syncr{\ar}{\hp}{\Theta}{x\sigma:B_1\wedgep B_2}, \Omega_1, \Omega_2}
			      {\syncr{\ar}{\hp}{\Theta}{x\sigma:B_1, \Omega_1}\quad\syncr{\ar}{\hp}{\Theta}{x\sigma:B_2, \Omega_2}}
			\]
			\[
			\infer[\veepf, i \in \{1,2\}]{\syncr{\ar}{\hp}{\Theta}{x\sigma:B_1\veep\kern -2pt B_2, \Omega}}{\syncr{\ar}{\hp}{\Theta}{x\sigma:B_i, \Omega}}
			\quad
			\infer[\diamondf]{\syncr{\ar \cup \{x\rel y\}}{\hp}{\Theta}{x y:\lozenge B, \Omega}}{\syncr{\ar \cup \{x\rel y\}}{\hp}{\Theta}{y:B, \Omega}}
			\]
	%
	%
			{\sc Identity rules}
			\[
			\infer[\initf]{\syncr{\ar}{\hp}{x:\neg B,\Theta}{x:B}}{}
			\qquad
			\]
	%
			{\sc Structural rules}
			\[
			\infer[\storef]{\asyncr{\ar}{\hp}{\Theta}{x:B,\Omega}}{\asyncr{\ar}{\hp}{\Theta,x:B}{\Omega}}
			\quad\
			\infer[\releasef]{\syncr{\ar}{\hp}{\Theta}{\Omega}}{\asyncr{\ar}{\hp}{\Theta}{\Omega'}}
			\quad\
			\infer[\decidef]{\asyncr{\ar}{\hp}{\Theta}{\cdot}}{\syncr{\ar}{\hp'}{\Theta}{\Omega^{}}}
			\]

			\bigskip

			In $\storef$, $B$ is a positive formula or a negative literal.\\
			In $\initf$, $B$ is a positive literal.\\
			In $\boxf$, $y$ is different from $x$ and does not occur in $\ar$ nor in $\Theta$.
			\\
			In $\decidef$, if $xy:A \in \Omega$ then $x:A \in \Theta$. Moreover, $\Omega$ contains only positive formulas of the form: $(i)$ $x\sigma:A$, where $A$ is not a $\lozenge$-formula and $x\in {\hp}$; or $(ii)$ $xy:A$ where $A$ is a $\lozenge$-formula, $x\rel y \in \ar$, $x\in {\hp}$.\\
			In $\releasef$, $\Omega$ contains no positive formulas and $\Omega' = \{x:A \mid x\sigma:A\in \Omega\}$. \\
			\caption{$\lmfstar$: a focused labeled framework for the modal logic $K$.}
			\label{fig:lmfstar}
			\end{figure}


\subsection{Emulation of modal proof systems}
	\label{sec:emul-theory}

	In order to emulate proofs given in other proof calculi by means of the focused framework $\lmfstar$, we need first of all to define a translation from the original modal language to the polarized one.

	 When translating a modal formula into a polarized one, we are often in a situation where we are interested in putting a delay in front of the formula only in the case when it is negative and not a literal. For that purpose, we define $\delp{A}$, where $A$ is a modal formula in negation normal form, to be $A$ if $A$ is a literal or a positive formula and $\delayop^+(A)$ otherwise. We then define the translation $\polos{.}$ as follows:
		\begin{center}
			\begin{tabular}{ccc@{\qquad\qquad}ccc}
				\small$\polos{P}$ &\small = &\small $P$ &
				\small$\polos{A \wedge B}$ &\small = &\small $\delp{\polos{A}} \wedgen \delp{\polos{B}}$\\
				\small$\polos{\neg P}$ &\small = &\small $\neg P$ &
				\small$\polos{A \vee B}$ &\small = &\small $\delp{\polos{A}} \veen \delp{\polos{B}}$\\
				$\polos{\square A}$ & \small= &\small $\square(\delp{\polos{A}})$ &
				\small$\polos{\lozenge {A}}$ &\small = &\small $\lozenge(\delayop^-(\delp{{\polos{A}}}))$
			\end{tabular}
		\end{center}

	In this translation, delays are used to ensure that only one connective of the original formula is processed along a given bipole of a focused derivation of the translated formula. This will be useful, in our proof checking procedure, in order to keep a tight connection between the original proof and the reconstructed one.

	Finally, in order to emulate existing calculi in $\lmfstar$, we need to give specialized versions of the rule $\decidef$.
	For $\lss$, we do it as follows:
	{$$
		\infer[\decide_{LS}]{\asyncr{\ar}{\lab}{\Theta}{\cdot}}{\syncr{\ar}{\lab}{\Theta}{x\sigma:A}}
		$$}
		where:
		\begin{itemize}
			\item $\lab$ denotes the set of all labels;
			\item if $A$ is a $\lozenge$ formula, then $\sigma$ is $y$ for some $x\rel y \in \ar$; otherwise, $\sigma$ is empty.
		\end{itemize}

		Given the similar nature of the approaches, in the case of the logic $K$, the same rule can be used also for emulating the systems $\pt$ and $\nss$ (for convenience, in the following we will use for the same rule also the names $\decide_{PT}$ and $\decide_{NS}$).

		For the system $\os$, we specialize instead the rule $\decidef$ as follows:
	{$$
	\infer[\decide_{OS}]{\asyncr{\ar}{\{x\}}{\Theta}{\cdot}}{\syncr{\ar}{\{y\}}{\Theta}{\Omega^{}}}
	$$}
		where (in addition to the general conditions of Fig.~\ref{fig:lmfstar}) we have that:
		\begin{enumerate}[label=(\arabic*)]
		\item if $x \neq y$, then:
			\begin{itemize}
					\item $x \rel y \in \ar$; and
					\item $\Omega$ is a multiset of formulas of the form $xy:\lozenge A$;
			\end{itemize}
		\item if $x = y$, then $\Omega = \{x:A\}$ for some formula $A$ that is not a $\lozenge$-formula.
		\end{enumerate}

		Intuitively, the specialization with respect to the general framework consists in: $(i)$ restricting the use of multi-focusing to $\lozenge$-formulas; $(ii)$ forcing such $\lozenge$-formulas to be labeled with the same future. This restriction is driven by the need for reproducing the behavior of the $\os$ modal rule $\krule$.

		Let $X$ range over $\{LS, PT, OS, NS\}$.
		We call $\lmf_{X}$ the system obtained from $\lmfstar$ by replacing the rule $\decidef$ with the rule $\decide_{X}$.
		The following adequacy result is proved by associating to each rule in $X$ a \emph{corresponding} sequence of bipoles in $\lmf_X$.
		We refer the reader to~\cite{MarMilVol16} for a more formal statement of the theorem as well as for its complete proof.

	\begin{theorem}\label{th:lmfstar}
		Let $X$ range over $\{LS, PT, OS, NS\}$. There exists a proof $\Pi$ of $A$ in the proof system $X$ iff there exists a proof $\Pi'$ of ${\asyncr{\emptyset}{\{x\}}{x:\delp{(\polos{A})}}{\cdot}}$, for any $x$, in $\lmf_{X}$. Moreover, for each application of a rule $r$ in $\Pi$ there is a sequence of bipoles in $\Pi'$ corresponding to $r$.
	\end{theorem}

	The result in Theorem~\ref{th:lmfstar} establishes a relation between the original calculi to be emulated and $\lmfstar$. Since our ultimate goal is to certify proofs in a kernel which consists in $\LKF$, we need to be able to relate $\lmfstar$ and $\LKF$ as well.
	First of all, based on the standard translation of Sec.~\ref{sec:std-translation}, we refine the translation above in order to consider also the translation of modalities into quantified formulas.

		Given a variable $x$, we define the translation $\tr{.}{x}$ from modal formulas in negation normal form into polarized first-order formulas as follows:
		\begin{center}
		\begin{tabular}{ccc@{\qquad\qquad}ccc}
			\small$\tr{P}{x}$ &\small = &\small $P(x)$ &
			\small$\tr{{A} \wedge {B}}{x}$ &\small = &\small $\delp{\tr{{A}}{x}} \wedgen \delp{\tr{{B}}{x}}$\\
			\small$\tr{\neg P}{x}$ &\small = &\small $\neg P(x)$ &
			\small$\tr{{A} \vee {B}}{x}$ &\small = &\small $\delp{\tr{{A}}{x}} \veen \delp{\tr{{B}}{x}}$\\
			\small	$\tr{\square {A}}{x}$ & \small= &\small $\forall y (\neg \relfo(x,y) \veen \delp{\tr{{A}}{y}})$
			 &
			\small$\tr{\lozenge {A}}{x}$ &\small = &\small $\exists y (\relfo(x,y) \wedgep \delayop^-(\delp{\tr{{A}}{y}}))$
		\end{tabular}
		\end{center}
	We observe that the translation of modalities also makes use of delays, in such a way that the processing of a modality in the labeled calculus corresponds to a bipole in $\LKF$, e.g., when in $\LKF$ we focus on a formula $\tr{\lozenge {A}}{x}$, the formula $A$ is delayed in such a way that it gets necessarily stored at the end of the bipole.
		Based on that, we define the translation $\trlab{.}$ from labeled
		formulas and relational atoms into polarized first-order formulas as
		$\trlab{x:A} = \tr{{A}}{x}$ and $\trlab{x\rel y} =\relfo(x,y)$.
	%
		%
		Predicates of the form $P(x)$ and $\relfo(x,y)$ are assigned positive polarity.
		This translation will be used for all the formalisms considered in the paper.

	It is easy to notice that each proof in $\lmfstar$ is also a proof of $\lmfm$ (just ignore the present of a sequent, as well as the present and future of formulas).
	Furthermore a proof in the multi-focused system $\lmfm$ can always be reproduced in $\lmf$, by breaking a multi-focused bipole into a chain of single-focused bipoles. Finally, in~\cite{MilVol15} it has been shown that a strict correspondence between proofs in $\lmf$ and proofs in $\LKF$ (restricted to the correspondence language) exists.
	This chain of correspondences allows us to state the following theoretical result, on which the adequacy of the implementation proposed in next section relies.

	\begin{theorem}\label{th:emulation}
		Let $X$ range over $\{LS, PT, OS, NS\}$. There exists a proof $\Pi$ of $A$ in the proof system $X$ iff there exists a proof $\Pi'$ of $\tr{A}{x}$ in $\LKF$, for any $x$. Moreover, for each application of a rule $r$ in $\Pi$ there is a sequence of bipoles in $\Pi'$ corresponding to $r$.
	\end{theorem}

\section{Certification of modal proofs}
\label{sec:cert}

This section describes the implementation of a general framework for the certification of modal proofs and shows
how this framework can be used in order to certify proofs from different proof systems.
We will rely here on the theoretical results of Sec.~\ref{sec:framework-aiml}.

The implementation discussed in this paper is freely accesible on Github \footnote{\url{https://github.com/proofcert/checkers/tree/dalefest}} or on Zenodo\footnote{\url{https://zenodo.org/record/1325924#.W2IWPHVfgWM}}. More information on the implementation
is given in Sec.~\ref{sec:examples}.



Foundational proof certificates (see Sec.~\ref{sec:checkers} for details)
form a rich language for the certification of any proof object. This flexibility stems from
connecting a trusted kernel with arbitrary $\lambda$Prolog programs (called FPC specifications).
The richness of the language, however, has the downside that defining a new set of FPC specifications is, in general, a complex
task -- it involves the encoding of the semantics of a system over another (represented by the kernel).
The complexity of supporting a new proof format is not unique to \pcert.
There are but a few general proof certification tools and
the effort to enable the certification of a particular proof system is non-trivial.

Our aim in this paper is to create a certification framework which enjoys generality and ease of use.
Taking, for example, the $\os$ system given in Fig.~\ref{fig:os}, Theorem \ref{th:emulation} has established the existence
of a functional transformation from ordinary sequent proofs into $\lmf_{OS}$, which is a restriction of $\lmfstar$.
In \cite{MilVol15}, the existence of a functional transformation from $\lmf$ proofs into \LKF\ was shown to exist.

A simple approach to the certification of ordinary sequent proofs would then be the direct translation of these proofs into \LKF.
%
Such a translation would amount to a soundness proof of the ordinary sequent calculus and would violate our two criteria mentioned above. Generalization would be violated since the translation would
target ordinary sequent proofs only. 
Moreover, since such translations are generally very complex and include many technical details that are hidden in the theoretical proofs, they are not easy to implement.

In this paper we present an approach based on the general proof checker presented in Sec.~\ref{sec:checkers}.
We attempt to encode the semantics of different proof systems using logical programs (predicates), a trusted kernel and proof guidance and search.
But, while defining the semantics using logical predicates is easier than using a functional translation, we are still left with the
complexity of defining these predicates. In addition, it seems that by writing the predicates for a particular calculus, we compromise on generality.

A way to amend the generalization problem of the two approaches above is to consider a framework in which many different proofs can
be certified. Such general frameworks are often cumbersome to use. Therefore, in order to make the certification as easy as possible,
we would like to require the framework semantics
to be as close to the semantics of the different proof calculi as possible.

The two properties stated above seem contradictory to each other. Being general and supporting different proof calculi and formats necessarily
mean making it harder to implement any specific calculus and format.
We try to circumvent this problem by introducing different layers in our framework. Some layers will be very general but harder to use
while others will be simpler but would not be able to support as many formats. In addition, the layers will be build on top of each
other in such a way that using an upper layer necessarily means using also a lower one. As we will see, the simplest and lowest layer
will be \LKF.  The remaining layers correspond to the systems which were introduced
in Sec.~\ref{sec:lmf} and Sec.~\ref{sec:framework}.

It should be noted that, as long as we always use the same lower level kernel, at no point trust is being compromised. Both the functional and the logical approaches are based on the reconstruction
of a proof in the classical first-order sequent calculus. If such a proof is constructed for a certain (translation of the) original formula,
we are assured that the formula is valid, up to the correctness of the first-order certifier as well as of the adequacy of the translation.

A simple layered approach would consist in using a separate kernel for
each layer, but that would compromise some of the trust we can place in the proof certifier
as well as violate the universality of the certification process, since certifications over different kernels
cannot be combined into one, foundational, proof.
For this reason, we will stick to one concrete kernel (\LKF) and will simulate the different layers on top of it.

\subsection{Introducing the framework}

In this section we present an implementation of a proof certification framework based on the general proof checker from Sec.~\ref{sec:checkers}
and which uses \LKF\ as its sole kernel. As mentioned above, such an approach enjoys the highest amount of trust.

The layers of our architecture will correspond to the implementation of systems that were described in the previous sections and that we briefly recall here:
\begin{enumerate}[label=(\arabic*)]
  \item {\LKF\ : a focused sequent calculus for first-order classical logic; from~\cite{LiaMil09} -- see Sec.~\ref{sec:lkf};}
  \item {$\lmf$\ - a focused labeled sequent calculus for the modal logic $K$; from~\cite{MilVol15} -- see Sec.~\ref{sec:lmf}};
  \item {$\lmfm$\ - a multi-focused variant of $\lmf$ -- see Sec.~\ref{sec:lmf}};
  \item {$\lmfs$\ - a framework, based on $\lmfm$, for the emulation of modal calculi; from~\cite{MarMilVol16} -- see Sec.~\ref{sec:framework}}.
\end{enumerate}
Given, for example, the proof evidence for the $\os$ proof in Fig.~\ref{fig:osex}, such a layered architecture allows us to certify the evidence over $\LKF$, while defining the FPC over another layer, which is closer
to the semantics of ordinary sequents.

\begin{figure}

\begin{prooftree}
\AxiomC{}
\UnaryInfC{$\vdash P,\neg P, Q$}
\AxiomC{}
\UnaryInfC{$\vdash \neg Q,\neg P, Q$}
\BinaryInfC{$\vdash P\wedge\neg Q, \neg P, Q$}
\UnaryInfC{$\vdash\lozenge(P\wedge\neg Q), \lozenge\neg P, \square Q$}
\UnaryInfC{$\vdash\lozenge(P\wedge\neg Q) , \lozenge\neg P \vee \square Q$}
\UnaryInfC{$\vdash\lozenge(P\wedge\neg Q) \vee (\lozenge\neg P \vee \square Q)$}
\end{prooftree}

  \caption{An $\os$ proof of the axiom $K$.}
  \label{fig:osex}
\end{figure}

First, let us examine the information in the proof. The information contained in each inference step of an $\os$ proof can be summarized as follows:
\begin{itemize}
  \item the (main) formula occurrence to which the inference is applied (in case of $\boxos$, we consider this to be the $\square$-formula introduced; in case of $\initos$, we consider this to be the positive literal in the couple of complementary literals);
  \item a possible additional list of formula occurrences:
    \begin{itemize}
      \item in case of the rule $\boxos$, all the $\lozenge$-formula occurrences that are introduced by the rule;
      \item in case of the rule $\initos$, the complementary negative literal.
    \end{itemize}
\end{itemize}

Therefore, an adequate tree-shaped proof evidence for the above $\os$ proof is the one shown in Fig.~\ref{fig:c1}.

\begin{figure}
\begin{itemize}
  \item $\lozenge(P\wedge\neg Q) \vee (\lozenge\neg P \vee \square Q)$
  \item $\lozenge\neg P \vee \square Q$
  \item $\square Q$ with additional information $\lozenge(P\wedge\neg Q)$ and $\lozenge\neg P$
  \item $P \wedge \neg Q$
  \begin{itemize}
    \item $P$ and the additional information $\neg P$
    \item $Q$ and the additional information $\neg Q$
  \end{itemize}
\end{itemize}
 \caption{The $OS$ proof evidence for a proof of the axiom $K$.}
  \label{fig:c1}
\end{figure}

In the next parts of this section, we will use the above example in order to discuss the implementation details of different
aspects of the framework.

\subsection{Formula indices}

Our first challenge is to be able to refer to specific formula occurrences inside a proof whose conclusion contains a single formula $A$.
The role of the indices will be to identify a specific subformula of $A$.
For example, given an index $I$ for a conjunctive formula, the index of the left conjunct can be defined as $\lft(I)$.

\begin{definition}[Basic indexing]
  \label{def:bind}
Given a formula $F$ which has an index $I$, we define the following indices for the subformulas of $F$:
\begin{itemize}
  \item if $F = A \wedge B$ or $F = A \vee B$, we assign the index $\lft(I)$ to $A$ and $\rgt(I)$ to $B$;
  \item if $F = \square A$ or $F = \lozenge A$, we assign the index $\lft(I)$ to $A$.
\end{itemize}
\end{definition}

Using the above basic indexing scheme and denoting the index of the theorem to prove by $\emptyset$,
the proof from Fig.~\ref{fig:osex} will be represented by the evidence in Fig.~\ref{fig:c2}.

\begin{figure}
\begin{itemize}
  \item $\emptyset$
  \item $\rgt(\emptyset)$
  \item $\rgt(\rgt(\emptyset))$ with the additional information $\lft(\emptyset)$ and $\lft(\rgt(\emptyset))$
  \item $\lft(\lft(\emptyset))$
  \begin{itemize}
    \item $\lft(\lft(\lft(\emptyset)))$ with the additional information $\lft(\lft(\rgt(\emptyset)))$
    \item $\lft(\rgt(\rgt(\emptyset)))$ with the additional information $\lft(\rgt(\lft(\emptyset)))$
  \end{itemize}
\end{itemize}
 \caption{$OS$ proof evidence using basic indexing.}
  \label{fig:c2}
\end{figure}

While sufficient for an ordinary sequent calculus for the system K, this indexing mechanism falls short for most other systems
and calculi. The reason for that is represented by the implicit contractions of formulas which take place in such systems (for example, in $LS$, in the introduction rule for $\lozenge$).
In our representation of proofs, this amounts to the need for indexing differently possible distinct occurrences of direct subformulas of a given $\lozenge$-formula. In order to distinguish between them, we define a correspondence between $\lozenge$-formulas and $\Box$-formulas inside a given proof.
%
This will also allow us to capture the idea behind labels and to therefore omit explicit label information in our framework.\footnote{In fact, in a (single-sided) labeled system, the correspondence would rather be between a $\Diamond$ introduction rule application and a specific label. However, at least in the case of the logic $K$, the only way to introduce (bottom-up) a new label is by means of a $\square$ introduction rule. It is thus possible to identify each label with the $\Box$-formula that introduces it.}


\begin{definition}[Modal correspondence for \lmfs]
  \label{df:cor}
  Let $\Pi$ be an $\lmfs$ proof. We say that a labeled formula $x: \Diamond A$ \emph{corresponds in  $\Pi$} to a labeled formula $y: \Box B$ iff:
  \begin{itemize}
    \item there is a $\Box_F$ rule application in $\Pi$ whose conclusion is a sequent containing $y: \Box B$ and whose premise is a sequent containing the formula $z: B$; and
\item there is a $\Diamond_F$ rule application in $\Pi$ whose conclusion is a sequent containing the formula $x: \Diamond A$ and whose premise is a sequent containing the formula $z: A$.
\end{itemize}
\end{definition}

This definition can be extended to non-labeled systems such as $OS$ and $NS$. For example,
upon the application of the $OS$ inference rule $\boxos$, all $\lozenge$-formulas in the lower sequent corresponds to the
$\square$-formula. Similarly for nested sequents, $\Diamond$ inference rules result in formulas being added to nested sequents which
are associated with $\Box$-formulas.

\begin{definition}[Modal correspondence for ordinary sequents]
  \label{df:cor2}
  Let $\Pi$ be an $\os$ proof. A formula $\Diamond A$ \emph{corresponds in $\Pi$} to a formula $\Box B$ iff:
  \begin{itemize}
    \item the conclusion of a $\boxos$ rule application in $\Pi$ contains both formulas.
\end{itemize}
\end{definition}

\begin{definition}[Modal correspondence for nested sequents]
  \label{df:cor3}
  Let $\Pi$ be an $NS$ proof. A formula $\Diamond A$ \emph{corresponds in $\Pi$} to a formula $\Box B$ iff:
  \begin{itemize}
    \item there is a $\Box_{NS}$ rule application in $\Pi$ whose conclusion contains  $\Gcon{\square B}$ and whose premise contains $\Gcon{[B]}$; and
    \item there is a $\Diamond_{NS}$ rule application in $\Pi$ whose conclusion contains $\Gcon{\Diamond A,[\mathcal{M}]}$ and whose premise contains $\Gcon{[A,\mathcal{M}]}$, where $\mathcal{M}$
      contains $B$.
\end{itemize}
\end{definition}

%

We omit the definitions for $\lss$ and $\pt$, which can be easily inferred from the one given for $\lmfstar$ and are anyway not used in the examples of the paper.

Using modal correspondence, we can define the indices of $\lozenge$-formulas.

\begin{definition}[Indexing]
Given a formula $F$ which has an index $I$, we refine the definition of basic indexing (\ref{def:bind})
and replace the indexing of direct subformulas of $\lozenge$-formulas as follows:
\begin{itemize}
  \item if we apply the diamond inference rule to a formula $F = \lozenge A$ corresponding to a $\square$-formula at index
    $J$, we assign the index $\diaind(I,J)$ to $A$.
\end{itemize}
\end{definition}

Using the indexing mechanism above on our example, we get the evidence in Fig.~\ref{fig:c3}.

\begin{figure}
\begin{itemize}
  \item $\emptyset$
  \item $\rgt(\emptyset)$
  \item $\rgt(\rgt(\emptyset))$ with the additional information $\lft(\emptyset)$ and $\lft(\rgt(\emptyset))$
	\item $\diaind(\lft(\emptyset), \rgt(\rgt(\emptyset)))$
  \begin{itemize}
    \item $\lft(\diaind(\lft(\emptyset), \rgt(\rgt(\emptyset))))$ \\
      with the additional information $\diaind(\lft(\rgt(\emptyset)), \rgt(\rgt(\emptyset)))$
    \item $\lft(\rgt(\rgt(\emptyset)))$ \\
      with the additional information $\rgt(\diaind(\lft(\emptyset), \rgt(\rgt(\emptyset))))$
  \end{itemize}
\end{itemize}
 \caption{$OS$ proof evidence using indexing.}
  \label{fig:c3}
\end{figure}
\subsection{Layered architecture}

As mentioned in the introduction to this section, our aim is to implement a layered framework. On the one hand,
the upper you go, the more restricted system you reach in which it is easier to define complex semantics of proof calculi.
On the other hand, as you go down, the layer syntax and definition are simpler and might be a better fit for some proof calculi.

No matter which layer we are using, we still wish our certification to be over our most trusted kernel - \LKF. Let us consider the layer just above it
- \lmf. Given a proof evidence denoted in terms of the system of this layer, it is straightforward to define its FPC specification
over \LKF. When we consider one layer above - \lmfm, it becomes more complex to define it over \LKF\ but a relatively simpler matter
to define it in terms of \lmf.

\subsubsection{The \lmf\ system layer.}

In the previous examples, we have seen that modal proof evidences normally contain information about the worlds
associated with $\Box$ and $\Diamond$ inference rules.
Our first layer is capable, therefore, of accepting proof evidences which contain the following information:

\begin{enumerate}[label=(\arabic*)]
	\item at each step, on which formula we apply a rule of the $\lmf$ calculus;
	\item in the case of a $\lozenge$-formula,
    with respect to which label (or, equivalently, as explained in the previous section, with respect to which $\Box$-formula) we apply the rule;
	\item in the case of an initial  rule, with respect to which complementary literal we apply it.
\end{enumerate}

For this reason, we define the proof evidence of this layer
as a tree describing the original proof.
Each node is decorated by a pair containing: (i) the index of the formula on which a rule is applied, as explained in (1),
together with (ii) a (possibly null) further index carrying additional information, to be used in cases (2) and (3) above.
Formulas in the tree will drive the construction (bottom-up) of the \LKF derivation, in the sense that,
by starting from the root, at each step, the \LKF kernel will decide on the given formula and proceed,
constrained by properly defined clerks and experts, along a synchronous and an asynchronous phase.
The results in \cite{LibalV16} guarantee that at the end of a bipole, we will be in a situation which
is equivalent to that of the corresponding step in the original proof.

As described in item (2) above, if we are applying an $\exists$-rule in \LKF, then we need
further information specifying with respect to which term we apply the rule. This term can be found in the additional
index supplied.
Similarly, in the case of an initial (3), the additional information in the node will specify the index of the complementary literal.

This definition permits the usage of different types of nodes in the same tree, which will allow us to smoothly move between the layers.

Using these definitions, we can now denote our example from Fig.~\ref{fig:c3} in terms of the \lmf\ layer, as can be seen in Fig.~\ref{fig:c4}.
The corresponding $\lmf$ proof is shown in Fig.~\ref{fig:lmf-proof}. One can observe that each item in the proof evidence of Fig.~\ref{fig:c4} corresponds to a block in the derivation of Fig.~\ref{fig:lmf-proof} that starts, reading the proof bottom-up, with a $\decidek$ application and possibly uses, along the bipole, the additional information contained in the evidence.
Note that, for shortness, we use in this proof (as well as in other examples below) derived rules for $\delayop^{+}$ and $\delayop^{-}$ introduction. They can be easily derived from other rules in the calculi.

\begin{figure}
\begin{itemize}
  \item $\emptyset$
  \item $\rgt(\emptyset)$
  \item $\rgt(\rgt(\emptyset))$
  \item $\lft(\emptyset)$ with the additional information $\rgt(\rgt(\emptyset))$
  \item $\lft(\rgt(\emptyset))$ with the additional information $\rgt(\rgt(\emptyset))$
	\item $\diaind(\lft(\emptyset), \rgt(\rgt(\emptyset)))$
\begin{itemize}
    \item $\lft(\diaind(\lft(\emptyset), \rgt(\rgt(\emptyset))))$ \\
      with the additional information $\diaind(\lft(\rgt(\emptyset)), \rgt(\rgt(\emptyset)))$
    \item $\lft(\rgt(\rgt(\emptyset)))$ \\
      with the additional information $\rgt(\diaind(\lft(\emptyset), \rgt(\rgt(\emptyset))))$
  \end{itemize}
\end{itemize}
 \caption{The $\lmf$ proof evidence for a proof of the axiom $K$.}
  \label{fig:c4}
\end{figure}

\begin{figure}
\footnotesize
	\infer[\decidek]{\asyncr{\emptyset}{}{x:\varphi}{\cdot}}{\infer[\releasek]{\syncr{\emptyset}{}{x:\varphi}{x:\varphi}}{\infer[\veenk]{\asyncr{\emptyset}{}{x:\varphi}{x:\varphi}}{\infer[\storek]{\asyncr{\emptyset}{}{x:\varphi}{x:\lozenge(\negd{\posd {P \wedgen \neg Q}})\, , \, x:\posd{\lozenge \negd{\neg P} \veen \posd{\square Q}}}}{\infer[\storek]{\asyncr{\emptyset}{}{\Sigma \equiv \{x:\varphi\, , \, x:\lozenge(\negd{\posd{P \wedgen \neg Q}})\}}{x:\posd{\lozenge \negd{\neg P} \veen \posd{\square Q}}}}{\infer[\decidek]{\asyncr{\emptyset}{}{\Sigma_1 \equiv \Sigma \cup \{x:\posd{\lozenge \negd{\neg P} \veen \posd{\square Q}}\}}{\cdot}}{\infer[\delayop^{+}]{\syncr{\emptyset}{}{\Sigma_1}{x:\posd{\lozenge \negd{\neg P} \veen \posd{\square Q}}}}{\infer[\releasek]{\syncr{\emptyset}{}{\Sigma_1}{x:\lozenge \negd{\neg P} \veen \posd{\square Q}}}{\infer[\veenk]{\asyncr{\emptyset}{}{\Sigma_1}{x:\lozenge \negd{\neg P} \veen \posd{\square Q}}}{\infer[\storek]{\asyncr{\emptyset}{}{\Sigma_1}{x:\lozenge \negd{\neg P}\, , \, x: \posd{\square Q}}}{\infer[\storek]{\asyncr{\emptyset}{}{\Sigma_2 \equiv \Sigma_1 \cup \{x:\lozenge \negd{\neg P}\}}{x:\posd{\square Q}}}{\infer[\decidek]{\asyncr{\emptyset}{}{\Sigma_3 \equiv \Sigma_2 \cup \{ x:\posd{\square Q}\}}{\cdot}}{\infer[\delayop^{+}]{\syncr{\emptyset}{}{\Sigma_3}{x:\posd{\square Q}}}{\infer[\releasek]{\syncr{\emptyset}{}{\Sigma_3}{x:\square Q}}{\infer[\boxk]{\asyncr{\emptyset}{}{\Sigma_3}{x:\square Q}}{\infer[\storek]{\asyncr{\{x \rel y\}}{}{\Sigma_3}{y: Q}}{\infer[\decidek]{\asyncr{\{x \rel y\}}{}{\Sigma_4 \equiv \Sigma_3 \cup \{y: Q\}}{\cdot}}{\infer[\diamondk]{\syncr{\{x \rel y\}}{}{\Sigma_4}{x:\lozenge(\negd{\posd{P \wedgen \neg Q}})}}{\infer[\releasek]{\syncr{\{x \rel y\}}{}{\Sigma_4}{y:\negd{\posd{P \wedgen \neg Q}}}}{\infer[\delayop^{-}]{\asyncr{\{x \rel y\}}{}{\Sigma_4}{y:\negd{\posd{P \wedgen \neg Q}}}}{\infer[\storek]{\asyncr{\{x \rel y\}}{}{\Sigma_4}{y:\posd{P \wedgen \neg Q}}}{\infer[\decidek]{\asyncr{\{x \rel y\}}{}{\Sigma_5 \equiv \Sigma_4 \cup \{y:\posd{P \wedgen \neg Q}\}}{\cdot}}{\infer[\diamondk]{\syncr{\{x \rel y\}}{}{\Sigma_5}{x:\lozenge \negd{\neg P}}}{\infer[\releasek]{\syncr{\{x \rel y\}}{}{\Sigma_5}{y:\negd{\neg P}}}{\infer[\delayop^{-}]{\asyncr{\{x \rel y\}}{}{\Sigma_5}{y:\negd{\neg P}}}{\infer[\storek]{\asyncr{\{x \rel y\}}{}{\Sigma_5}{y:\neg P}}{\infer[\decidek]{\asyncr{\{x \rel y\}}{}{\Sigma_6 \equiv \Sigma_5 \cup \{y:\neg P\}}{\cdot}}{\infer[\delayop^{+}]{\syncr{\{x \rel y\}}{}{\Sigma_6}{y:\posd{P \wedgen \neg Q}}}{\infer[\releasek]{\syncr{\{x \rel y\}}{}{\Sigma_6}{y:P \wedgen \neg Q}}{\infer[\wedgenk]{\asyncr{\{x \rel y\}}{}{\Sigma_6}{y:P \wedgen \neg Q}}{\infer[\storek]{\asyncr{\{x \rel y\}}{}{\Sigma_6}{y:P}}{\infer[\decidek]{\asyncr{\{x \rel y\}}{}{\Sigma_7 \equiv \Sigma_6 \cup \{y:P\}}{\cdot}}{\infer[\initk]{\syncr{\{x \rel y\}}{}{\Sigma_7}{y:P}}{}}} & \infer[\storek]{\asyncr{\{x \rel y\}}{}{\Sigma_6}{y:\neg Q}}{\infer[\decidek]{\asyncr{\{x \rel y\}}{}{\Sigma_8 \equiv \Sigma_6 \cup \{y:\neg Q\}}{\cdot}}{\infer[\initk]{\asyncr{\{x \rel y\}}{}{\Sigma_8}{y:Q}}{}}}}}}}}}}}}}}}}}}}}}}}}}}}}}}}}}
 \caption{The $\lmf$ proof corresponding to the proof evidence of Fig.~\ref{fig:c4}, where $\varphi \equiv \lozenge(\negd{\posd{P \wedgen \neg Q}}) \veen (\posd{\lozenge \negd{\neg P} \veen \posd{\square Q}})$ is a polarized version of the axiom $K$, obtained according to the translation in Sec.~\ref{sec:emul-theory}.}
  \label{fig:lmf-proof}
\end{figure}


The implementation\footnote{\texttt{src/fpc/modal/lmf-singlefoc.mod}} of the \lmf\ layer over the \LKF\ kernel in our system will mainly do the following:
\begin{itemize}
  \item use the index in each node of the tree to choose the right formula to decide on;
  \item use the extra information for the $\lozenge$-formula when applying an $\exists$ rule;
  \item use the extra information for the literals when applying an $\texttt{init}$ rule.
\end{itemize}

A simplified version (omitting some technical details) of a part of the implementation is given in Fig.~\ref{fig:lkfimp}.
Here a definition consists of a predicate name, e.g.,~\verb+orNeg_c+, and a list of arguments. The names of the predicates correspond to those in Fig.~\ref{fig:lkf}, which enriches $\LKF$ with clerk (ending with \verb+_c+) and expert (ending with \verb+_e+) predicates.

%
The role of the arguments is also illustrated in Fig.~\ref{fig:lkf}. Typically, the first and last arguments of each predicate are the ``input'' and ``output'' evidence of the inference rule, i.e.,~the evidence of the conclusion and the evidence of the premise, respectively. In some cases, further arguments are present, e.g.,~in the \verb+some_e+ predicate of the example, a second argument containing information about the witness term to be used in the $\exists$-introduction application. In the implementation, an evidence of the $\lmf$ layer is a term
named \verb+lmf_cert+ which has two arguments: $(i)$ a state (holding additional information, as shown below) and $(ii)$ a tree.
The tree is a regular inductively defined tree \verb+node I O List+ where \verb+List+ is a list of sub-trees of the current node
and \verb+I+ and \verb+O+ are indices indicating the formulas to decide on, etc.

The first definition,
for the negative disjunction, just skips the root node of the proof tree in the evidence. Since we are in
an asynchronous phase, the information in the evidence is not required. On the other hand, the information transmitted to the clerk from
the kernel -
the principal formula - is ignored by the clerk since it is not required later. We also remark on the existence
of a state variable. The state is being used in order to propagate some information between the rule applications.
It is not used in this FPC definition.

The second definition, for the universal quantifier, does also skip the current node, similarly to the previous definition.
It does need, though, to record the eigenvariable used and stores it in a mapping in the state. This mapping associates
the optional index of the node, i.e., the index of the corresponding $\Box$-formula, to the actual eigenvariable introduced by the kernel.
We store these values in the state using a list of pairs of indices and eigenvariables. Since the predicate cannot
intervene directly with the kernel in order to obtain the eigenvariable, we have defined the second argument of this predicate
as a function from eigenvariables to proof evidences. The kernel is then responsible for applying this function to the eigenvariable.

The last definition, which is an expert FPC definition, selects the previously stored eigenvariable
and returns it as the term witness. In order to find the correct eigenvariable, we check for the membership of a pair
of the known index and the required eigenvariable in the list which forms the state.


\begin{figure}
\begin{lstlisting}[
  basicstyle=\small\ttfamily
]

orNeg_c
  (lmf-cert State (node I O [H|T]))
  (lmf-cert State H).

all_c
  (lmf-cert (state M) (node I O [H|T]))
  (X\ lmf-cert (state [pair I X|M]) H).

some_e
  (lmf-cert (state M) (node I O [H|T]))
  X
  (lmf-cert (state M) H) :-
     member (pair O X) M.
\end{lstlisting}
  \caption{A simplified version of the implementation of three FPC definitions.}
\label{fig:lkfimp}
\end{figure}

According to this specification, which can be found in \cite{LibalV16},
each $\decide$ step is completely determined by the proof evidence.

\subsubsection{The \lmfm\ system layer.}

One can immediately see that the \lmf\ layer is not the most suitable for describing the semantics of ordinary sequents.
The reason is that the order in which $\lozenge$-formulas are decided on, which is explicit in \lmf, is not always relevant in the corresponding ordinary sequent proof.

We would like to have a layer which allows us to decide simultaneously on different formulas in the sequent. This is
obtained by multi-focusing.

The \lmfm\ layer allows us to simulate a multi-focusing step in the kernel (which is non-multifocused) and corresponds to the multi-focused version of \lmf\
defined in Sec.~\ref{sec:lmf}.
Our system will simulate multi-focusing by relating each inference with a number.
This number will force all inferences labeled the same to occur sequentially.
We observe that this does not simulate multi-focusing adequately in the general case; however, for the modal proof calculi considered in this paper and due to the the fact that we restrict to the logic
$K$, we are ensured that this simple mechanism
is enough for encoding multi-focusing in our case.
We note here that in order to support multi-focusing in logics other than K,
we would need to support a multi-focused version of \LKF\ as our kernel.

A proof evidence for our running example in \lmfm\ can be seen in Fig.~\ref{fig:c5}. The multi-focus value
which appears there is just an integer which is used to group simultaneous focusing. The corresponding $\lmfm$ proof is shown in Fig.~\ref{fig:lmfm-proof}. With respect to the $\lmf$ proof of Fig.~\ref{fig:lmf-proof}, we can notice that here we have a $\decide$ step in which the focus is put on two $\lozenge$-formulas at the same time; such formulas correspond to those indices having the same multi-focus value in Fig.~\ref{fig:c5}.

\begin{figure}
\begin{itemize}
  \item $\emptyset$ with the additional information multi-focus value 1
  \item $\rgt(\emptyset)$ with the additional information multi-focus value 2
  \item $\rgt(\rgt(\emptyset))$ with the additional information multi-focus value 3
  \item $\lft(\emptyset)$ with the additional information $\rgt(\rgt(\emptyset))$ and multi-focus value 4
  \item $\lft(\rgt(\emptyset))$ with the additional information $\rgt(\rgt(\emptyset))$ and multi-focus value 4
\item $\diaind(\lft(\emptyset), \rgt(\rgt(\emptyset)))$ with the additional information multi-focus value 5
  \begin{itemize}
    \item $\lft(\diaind(\lft(\emptyset), \rgt(\rgt(\emptyset))))$ \\
      with the additional information $\diaind(\lft(\rgt(\emptyset)), \rgt(\rgt(\emptyset)))$ and multi-focus value 6
    \item $\lft(\rgt(\rgt(\emptyset)))$ \\
      with the additional information $\rgt(\diaind(\lft(\emptyset), \rgt(\rgt(\emptyset))))$ and multi-focus value 7
  \end{itemize}
\end{itemize}
 \caption{The $\lmfm$ proof evidence for a proof of the axiom $K$.}
  \label{fig:c5}
\end{figure}

\begin{figure}
\footnotesize
	\infer[\decidek]{\asyncr{\emptyset}{}{x:\varphi}{\cdot}}{\infer[\releasek]{\syncr{\emptyset}{}{x:\varphi}{x:\varphi}}{\infer[\veenk]{\asyncr{\emptyset}{}{x:\varphi}{x:\varphi}}{\infer[\storek]{\asyncr{\emptyset}{}{x:\varphi}{x:\lozenge(\negd{\posd {P \wedgen \neg Q}})\, , \, x:\posd{\lozenge \negd{\neg P} \veen \posd{\square Q}}}}{\infer[\storek]{\asyncr{\emptyset}{}{\Sigma \equiv \{x:\varphi\, , \, x:\lozenge(\negd{\posd{P \wedgen \neg Q}})\}}{x:\posd{\lozenge \negd{\neg P} \veen \posd{\square Q}}}}{\infer[\decidek]{\asyncr{\emptyset}{}{\Sigma_1 \equiv \Sigma \cup \{x:\posd{\lozenge \negd{\neg P} \veen \posd{\square Q}}\}}{\cdot}}{\infer[\delayop^{+}]{\syncr{\emptyset}{}{\Sigma_1}{x:\posd{\lozenge \negd{\neg P} \veen \posd{\square Q}}}}{\infer[\releasek]{\syncr{\emptyset}{}{\Sigma_1}{x:\lozenge \negd{\neg P} \veen \posd{\square Q}}}{\infer[\veenk]{\asyncr{\emptyset}{}{\Sigma_1}{x:\lozenge \negd{\neg P} \veen \posd{\square Q}}}{\infer[\storek]{\asyncr{\emptyset}{}{\Sigma_1}{x:\lozenge \negd{\neg P}\, , \, x: \posd{\square Q}}}{\infer[\storek]{\asyncr{\emptyset}{}{\Sigma_2 \equiv \Sigma_1 \cup \{x:\lozenge \negd{\neg P}\}}{x:\posd{\square Q}}}{\infer[\decidek]{\asyncr{\emptyset}{}{\Sigma_3 \equiv \Sigma_2 \cup \{ x:\posd{\square Q}\}}{\cdot}}{\infer[\delayop^{+}]{\syncr{\emptyset}{}{\Sigma_3}{x:\posd{\square Q}}}{\infer[\releasek]{\syncr{\emptyset}{}{\Sigma_3}{x:\square Q}}{\infer[\boxk]{\asyncr{\emptyset}{}{\Sigma_3}{x:\square Q}}{\infer[\storek]{\asyncr{\{x \rel y\}}{}{\Sigma_3}{y: Q}}{\infer[\decidek]{\asyncr{\{x \rel y\}}{}{\Sigma_4 \equiv \Sigma_3 \cup \{y: Q\}}{\cdot}}{\infer[\diamondk]{\syncr{\{x \rel y\}}{}{\Sigma_4}{x:\lozenge(\negd{\posd{P \wedgen \neg Q}}), x:\lozenge \negd{\neg P}}}{\infer[\diamondk]{\syncr{\{x \rel y\}}{}{\Sigma_4}{y:\negd{\posd{P \wedgen \neg Q}}, x:\lozenge \negd{\neg P}}}{\infer[\releasek]{\syncr{\{x \rel y\}}{}{\Sigma_4}{y:\negd{\posd{P \wedgen \neg Q}}, y:\negd{\neg P}}}{\infer[\delayop^{-}]{\asyncr{\{x \rel y\}}{}{\Sigma_4}{y:\negd{\posd{P \wedgen \neg Q}}, y:\negd{\neg P}}}{\infer[\storek]{\asyncr{\{x \rel y\}}{}{\Sigma_4}{y:\posd{P \wedgen \neg Q}, y:\negd{\neg P}}}{\infer[\delayop^{-}]{\asyncr{\{x \rel y\}}{}{\Sigma_5 \equiv \Sigma_4 \cup \{y:\posd{P \wedgen \neg Q}\}}{y:\negd{\neg P}}}{\infer[\storek]{\asyncr{\{x \rel y\}}{}{\Sigma_5}{y:\neg P}}{\infer[\decidek]{\asyncr{\{x \rel y\}}{}{\Sigma_6 \equiv \Sigma_5 \cup \{y:\neg P\}}{\cdot}}{\infer[\delayop^{+}]{\syncr{\{x \rel y\}}{}{\Sigma_6}{y:\posd{P \wedgen \neg Q}}}{\infer[\releasek]{\syncr{\{x \rel y\}}{}{\Sigma_6}{y:P \wedgen \neg Q}}{\infer[\wedgenk]{\asyncr{\{x \rel y\}}{}{\Sigma_6}{y:P \wedgen \neg Q}}{\infer[\storek]{\asyncr{\{x \rel y\}}{}{\Sigma_6}{y:P}}{\infer[\decidek]{\asyncr{\{x \rel y\}}{}{\Sigma_7 \equiv \Sigma_6 \cup \{y:P\}}{\cdot}}{\infer[\initk]{\syncr{\{x \rel y\}}{}{\Sigma_7}{y:P}}{}}} & \infer[\storek]{\asyncr{\{x \rel y\}}{}{\Sigma_6}{y:\neg Q}}{\infer[\decidek]{\asyncr{\{x \rel y\}}{}{\Sigma_8 \equiv \Sigma_6 \cup \{y:\neg Q\}}{\cdot}}{\infer[\initk]{\asyncr{\{x \rel y\}}{}{\Sigma_8}{y:Q}}{}}}}}}}}}}}}}}}}}}}}}}}}}}}}}}}
 \caption{The $\lmfm$ proof corresponding to the proof evidence of Fig.~\ref{fig:c5}, where $\varphi \equiv \lozenge(\negd{\posd{P \wedgen \neg Q}}) \veen (\posd{\lozenge \negd{\neg P} \veen \posd{\square Q}})$ is a polarized version of the axiom $K$, obtained according to the translation in Sec.~\ref{sec:emul-theory}.}
  \label{fig:lmfm-proof}
\end{figure}


\subsubsection{The \lmfs\ system layer.}

The most expressive layer is $\lmfs$.
This layer extends the previous one with information about worlds which are currently active (the present)
and the possible futures of formulas.

Going back to our running example, we see that we still cannot simulate properly the semantics of ordinary sequent calculus.
There is no mechanism in our \lmfm\ layer which enforces all $\lozenge$-formulas to ``go'' to the same world (the one introduced by the corresponding $\Box$-formula). Essentially,
we want to forbid proof evidences where $\Diamond$-formulas belonging to the same multi-focusing step correspond to different $\Box$-formulas, because we know that in such a case our kernel would not be simulating an $\os$ proof.
We can impose additional restrictions based on the tools from Sec.~\ref{sec:framework}, which will ensure such cases cannot happen.

Our running example proof evidence will now look like the one in Fig.~\ref{fig:c6}. The nodes of a proof evidence, as defined in Sec.~\ref{sec:framework},
contain, in addition to the information required in the previous layer, also information about the new present of the sequent and future of formulas, denoted by labels\footnote{We remark that, for simplicity and since it is equivalent in the case of the systems considered, in the implementation we attach the information concerning the future to the whole sequent, rather than to single formulas as described in Sec.~\ref{sec:framework}.}.
Essentially, the new features of this layer help us further restrict the proof search over the previous layer by giving us the ability to avoid applying the $\decide$ and $\Diamond$-introduction rules in some cases.
In Fig.~\ref{fig:lmfstar-proof}, we show the $\lmfstar$ proof corresponding to the proof evidence of Fig.~\ref{fig:c6}. The proof has the same structure as the one in Fig.~\ref{fig:lmfm-proof} and the way the additional ``decorations'' (denoting the present and the future) are used, respect the restrictions that characterize ordinary sequents and make it indeed an $\lmf_{OS}$ proof.

\begin{figure}
\begin{itemize}
  \item $\emptyset$ with the additional information multi-focus value 1, present $\{\emptyset\}$ and an empty future
  \item $\rgt(\emptyset)$ with the additional information multi-focus value 2, present $\{\emptyset\}$ and an empty future
  \item $\rgt(\rgt(\emptyset))$ with the additional information multi-focus value 3, present $\{\emptyset\}$ and an empty future
  \item $\lft(\emptyset)$ with the additional information $\rgt(\rgt(\emptyset))$, multi-focus value 4, present $\{\rgt(\rgt(\emptyset))\}$ and future $\rgt(\rgt(\emptyset))$
  \item $\lft(\rgt(\emptyset))$ with the additional information $\rgt(\rgt(\emptyset))$, multi-focus value 4, present $\{\rgt(\rgt(\emptyset))\}$ and future $\rgt(\rgt(\emptyset))$
\item $\diaind(\lft(\emptyset), \rgt(\rgt(\emptyset)))$ with the additional information multi-focus value 5, present $\{\rgt(\rgt(\emptyset))\}$ and an empty future
  \begin{itemize}
    \item $\lft(\diaind(\lft(\emptyset), \rgt(\rgt(\emptyset))))$ \\
      with the additional information $\diaind(\lft(\rgt(\emptyset)), \rgt(\rgt(\emptyset)))$, multi-focus value 6, present $\{\rgt(\rgt(\emptyset))\}$ and an empty future
    \item $\lft(\rgt(\rgt(\emptyset)))$ \\
      with the additional information $\rgt(\diaind(\lft(\emptyset), \rgt(\rgt(\emptyset))))$ and multi-focus value 7, present $\{\rgt(\rgt(\emptyset))\}$ and an empty future
  \end{itemize}
\end{itemize}
 \caption{The $\lmfstar$ proof evidence for a proof of the axiom $K$.}
  \label{fig:c6}
\end{figure}

\begin{figure}
\footnotesize
	\infer[\decidek]{\asyncr{\emptyset}{\{x\}}{x:\varphi}{\cdot}}{\infer[\releasek]{\syncr{\emptyset}{\{x\}}{x:\varphi}{x:\varphi}}{\infer[\veenk]{\asyncr{\emptyset}{\{x\}}{x:\varphi}{x:\varphi}}{\infer[\storek]{\asyncr{\emptyset}{\{x\}}{x:\varphi}{x:\lozenge(\negd{\posd {P \wedgen \neg Q}})\, , \, x:\posd{\lozenge \negd{\neg P} \veen \posd{\square Q}}}}{\infer[\storek]{\asyncr{\emptyset}{\{x\}}{\Sigma \equiv \{x:\varphi\, , \, x:\lozenge(\negd{\posd{P \wedgen \neg Q}})\}}{x:\posd{\lozenge \negd{\neg P} \veen \posd{\square Q}}}}{\infer[\decidek]{\asyncr{\emptyset}{\{x\}}{\Sigma_1 \equiv \Sigma \cup \{x:\posd{\lozenge \negd{\neg P} \veen \posd{\square Q}}\}}{\cdot}}{\infer[\delayop^{+}]{\syncr{\emptyset}{\{x\}}{\Sigma_1}{x:\posd{\lozenge \negd{\neg P} \veen \posd{\square Q}}}}{\infer[\releasek]{\syncr{\emptyset}{\{x\}}{\Sigma_1}{x:\lozenge \negd{\neg P} \veen \posd{\square Q}}}{\infer[\veenk]{\asyncr{\emptyset}{\{x\}}{\Sigma_1}{x:\lozenge \negd{\neg P} \veen \posd{\square Q}}}{\infer[\storek]{\asyncr{\emptyset}{\{x\}}{\Sigma_1}{x:\lozenge \negd{\neg P}\, , \, x: \posd{\square Q}}}{\infer[\storek]{\asyncr{\emptyset}{\{x\}}{\Sigma_2 \equiv \Sigma_1 \cup \{x:\lozenge \negd{\neg P}\}}{x:\posd{\square Q}}}{\infer[\decidek]{\asyncr{\emptyset}{\{x\}}{\Sigma_3 \equiv \Sigma_2 \cup \{ x:\posd{\square Q}\}}{\cdot}}{\infer[\delayop^{+}]{\syncr{\emptyset}{\{x\}}{\Sigma_3}{x:\posd{\square Q}}}{\infer[\releasek]{\syncr{\emptyset}{\{x\}}{\Sigma_3}{x:\square Q}}{\infer[\boxk]{\asyncr{\emptyset}{\{x\}}{\Sigma_3}{x:\square Q}}{\infer[\storek]{\asyncr{\{x \rel y\}}{\{x\}}{\Sigma_3}{y: Q}}{\infer[\decidek]{\asyncr{\{x \rel y\}}{\{x\}}{\Sigma_4 \equiv \Sigma_3 \cup \{y: Q\}}{\cdot}}{\infer[\diamondk]{\syncr{\{x \rel y\}}{\{y\}}{\Sigma_4}{xy:\lozenge(\negd{\posd{P \wedgen \neg Q}}), xy:\lozenge \negd{\neg P}}}{\infer[\diamondk]{\syncr{\{x \rel y\}}{\{y\}}{\Sigma_4}{y:\negd{\posd{P \wedgen \neg Q}}, xy:\lozenge \negd{\neg P}}}{\infer[\releasek]{\syncr{\{x \rel y\}}{\{y\}}{\Sigma_4}{y:\negd{\posd{P \wedgen \neg Q}}, y:\negd{\neg P}}}{\infer[\delayop^{-}]{\asyncr{\{x \rel y\}}{\{y\}}{\Sigma_4}{y:\negd{\posd{P \wedgen \neg Q}}, y:\negd{\neg P}}}{\infer[\storek]{\asyncr{\{x \rel y\}}{\{y\}}{\Sigma_4}{y:\posd{P \wedgen \neg Q}, y:\negd{\neg P}}}{\infer[\delayop^{-}]{\asyncr{\{x \rel y\}}{\{y\}}{\Sigma_5 \equiv \Sigma_4 \cup \{y:\posd{P \wedgen \neg Q}\}}{y:\negd{\neg P}}}{\infer[\storek]{\asyncr{\{x \rel y\}}{\{y\}}{\Sigma_5}{y:\neg P}}{\infer[\decidek]{\asyncr{\{x \rel y\}}{\{y\}}{\Sigma_6 \equiv \Sigma_5 \cup \{y:\neg P\}}{\cdot}}{\infer[\delayop^{+}]{\syncr{\{x \rel y\}}{\{y\}}{\Sigma_6}{y:\posd{P \wedgen \neg Q}}}{\infer[\releasek]{\syncr{\{x \rel y\}}{\{y\}}{\Sigma_6}{y:P \wedgen \neg Q}}{\infer[\wedgenk]{\asyncr{\{x \rel y\}}{\{y\}}{\Sigma_6}{y:P \wedgen \neg Q}}{\infer[\storek]{\asyncr{\{x \rel y\}}{\{y\}}{\Sigma_6}{y:P}}{\infer[\decidek]{\asyncr{\{x \rel y\}}{\{y\}}{\Sigma_7 \equiv \Sigma_6 \cup \{y:P\}}{\cdot}}{\infer[\initk]{\syncr{\{x \rel y\}}{\{y\}}{\Sigma_7}{y:P}}{}}} & \infer[\storek]{\asyncr{\{x \rel y\}}{\{y\}}{\Sigma_6}{y:\neg Q}}{\infer[\decidek]{\asyncr{\{x \rel y\}}{\{y\}}{\Sigma_8 \equiv \Sigma_6 \cup \{y:\neg Q\}}{\cdot}}{\infer[\initk]{\asyncr{\{x \rel y\}}{\{y\}}{\Sigma_8}{y:Q}}{}}}}}}}}}}}}}}}}}}}}}}}}}}}}}}}
 \caption{The $\lmfstar$ proof corresponding to the proof evidence of Fig.~\ref{fig:c6}, where $\varphi \equiv \lozenge(\negd{\posd{P \wedgen \neg Q}}) \veen (\posd{\lozenge \negd{\neg P} \veen \posd{\square Q}})$ is a polarized version of the axiom $K$, obtained according to the translation in Sec.~\ref{sec:emul-theory}.}
  \label{fig:lmfstar-proof}
\end{figure}


\subsection{Polymorphic proof evidence}
\label{sec:poly}

As we mentioned in previous sections, one of the goals of our framework is ease of use. When implementing the integration of specific modal systems,
we would like to use the layer which is closest to the semantics of the system, e.g., the $\lmfs$ layer in the case of $\os$.
When considering the proof evidences of these two systems (figures \ref{fig:c3} and \ref{fig:c6}),
we notice that they are not that similar.

In this section we present the first approach for amending this problem -- ease of use while preserving trust and universality -- which we call \emph{polymorphic proof evidence}.
The name stems from the fact that in order to use a layer and be able to certificate over the \LKF\ kernel, we consider different proof
evidences as being polymorphic, i.e., belonging to different proof calculi and layers.
%

In the example of $\os$, by polymorphic proof evidence, we mean that these proof evidences, from the implementation
point of view, are both ordinary sequent proofs and \lmfs\ proofs. In a similar way, we can define an \lmfs\ proof
as being both an \lmfs\ proof and an \lmfm\ one. By following this approach, we will obtain that the ordinary proof evidence, being
built on top of our topmost layer, is a proof evidence of all mentioned systems.

From the certification point of view, this will allow us to certify the $\os$ proof evidence ``out of the box'' by just considering
it as an \lmfs\ evidence. From the implementation point of view, we need to be able to define the evidence as
polymorphic. $\lambda$Prolog, being a logic programming language, does not support polymorphism on the object level (it does, though,
support type polymorphism similar to the one in functional programming languages).

We overcome that by using programs which translates the evidences across layers.

The first set of programs are based on the fact that each layer is built on top of the one below it.
We will include the proof evidence of the lower layer within the proof evidence of the upper one. Polymorphism will be obtained
by considering, in each layer, only the relevant component.

The second is based on the fact that a proof evidence, like that for an ordinary sequent proof, can be denoted in terms of
the proof evidence expected in one of the layers. Polymorphism is obtained via a logical program which translates the evidence back and forth between the original proof
and the one of the relevant layer.

By explicitly defining these small programs, we allow users to certify proof evidence, whose semantics are defined using one of the layers,
on top of the \LKF\ kernel.

An example of using a translation between an \lmfs\ and \lmfm\ proof evidences is given in Fig.~\ref{fig:smtns}.

A program which enables this ``polymorphic" behavior for ordinary sequents over our framework is given in Fig.~\ref{fig:ostns}.
In order to apply the FPC specification of the \lmfs\ layer, we convert the proof evidence to the one expected by the \lmfs\ layer
and recursively apply the predicate. The logic programming mechanism will use the transformed proof evidence in order to locate the proper predicate to apply. We can, therefore, simulate polymorphism over the proof evidences.

We can see that the essential information about the index
of the formula a rule is applied to, as well as the optional additional index (denoted \texttt{I} and \texttt{OI} in the figure)
are copied between the evidences. The information which is not part of the $OS$ evidence - the multi-focus index, the future and the
set of presents - is being stored in a state like data structure and is copied to the \lmfs\ evidence. The state of the $OS$
evidence is initially empty and is being updated by the $OS$ FPC specification.

Here we have shown the relatively simple (abstraction over the) definition of the \texttt{orNeg\_c} FPC specification.
The one for \texttt{allNeg\_c},
for example, includes a simulation of the application of $\Diamond$ inference rules.

\begin{figure}
\begin{lstlisting}[
  basicstyle=\small\ttfamily
]

star_to_multi-foc (star_cert S (multi-foc_cert M))
  (multi-foc_cert M) S.

multi-foc_to_star (multi-foc_cert M) S
  (star_cert S (multi-foc_cert M)).

orNeg_c Cert Form Cert' :-
 star_to_multi-foc Cert Cert-m S,
 % call matching a definition in the multi-foc layer
 orNeg_c Cert-m Form Cert-m',
 multi-foc_to_star Cert-m' S Cert'.
\end{lstlisting}
  \caption{Proof evidence transformation between two layers.}
\label{fig:smtns}
\end{figure}

\begin{figure}
\begin{lstlisting}[
  basicstyle=\small\ttfamily
]

ordinary_to_star
  (ordinary_cert (ordinary_state H F M) I OI)
  (star_cert H F (multi-foc_cert M (single_cert I OI))).

star_to_ordinary
  (star_cert H F (multi-foc_cert M (single_cert I OI)))
  (ordinary_cert (ordinary_state H F M) I OI).

orNeg_c Cert Form Cert' :-
 ordinary_to_star Cert Cert-s,
 orNeg_c Cert-s Form Cert-s',
 star_to_ordinary Cert-s' Cert'.
\end{lstlisting}
  \caption{Proof evidence transformation between $OS$ and $\lmfs$.}
\label{fig:ostns}
\end{figure}

\subsection{Certification of different proof formats}

Given the different layers in the proof system defined in the previous section, we can easily
write FPC specifications for different popular proof formats.

The process is always the same. The FPC specifications translate the evidence into the evidence of a particular layer,
as explained in Sec.~\ref{sec:poly}.

In the next sections we describe in more detail how the framework is used in order to support specific proof formats. In all cases, in order to perform the proof reconstruction inside $\LKF$, the modal formula to be proved is translated according to the translation $\tr{.}{}$ of Sec.~\ref{sec:emul-theory}.

\subsubsection{Labeled sequents}

The treatment of labeled systems ($\lss$) \cite{Neg05} was already implemented in the previous version of \checkers, which is described in \cite{LibalV16}.
In order to get emulation of $\lss$, we require a very simple use of the framework $\lmfstar$, where at each sequent the present corresponds to the set of all the labels occurring in the proof, no use of multi-focusing is required and the future of a formula is set, in the case of $\lozenge$-formulas, to the index of the corresponding $\square$-formula. For simplicity, since this is enough in the case of $K$, in our implementation we rely on the lower layer \lmf. Please refer to~\cite{LibalV16}.

\subsubsection{Prefixed tableaux}

The popular $\pt$ proof format \cite{fitting1972tableau}, which is used by various automated theorem provers (for example \cite{Beckert97a}), is, in the case of $K$, very close to that of $\lss$. Therefore support for it can be obtained in a very similar way. Its implementation, which has been described in \cite{LibalV16}, also relies on $\lmf$ and mainly consists in inverting, with respect to $\lss$ the role of boxes and diamonds in the FPC and in letting tableau closure rules behave as sequent initial rules.
This inversion is related to the fact that, despite $\lss$, $\pt$ is a refutation method.

\subsubsection{Ordinary Sequents}
\label{sec:os}


As described in Sec.~\ref{sec:bg-ordinary}, ordinary sequent systems ($\os$) differ in several ways from the previous systems. First, they do not have labels and second, they treat both $\square$ and $\lozenge$-formulas inside a single inference rule. For these reasons, the case of ordinary sequents illustrates the use of the features of the framework $\lmfstar$ in a more significant way already for the logic $K$.

In particular, the modal rule, which applies to all $\lozenge$-formulas at once, can be emulated in our system by using multi-focusing.
In addition, the relationship between the modal operators can be used in order to restrict the futures allowed: given a modal rule, all the $\lozenge$-formulas occurring there are assigned the same future, which corresponds to the index of the only $\square$-formula.

The information required in an ordinary sequent proof evidence is therefore:
\begin{itemize}
  \item For each application of a $\Box_F$ inference rule, a list of indices of all affected $\Diamond$-formulas.
  \item For the application of the initial rule, the corresponding index of the complementary literal
\end{itemize}

The program which translates between $OS$ and $\lmfs$ proof evidence needs to compute from the above information the relevant
multi-focus indices as well as the present and future of sequents and formulas.

An ordinary sequent node contains its index as well as a list of indices. This list is empty for all inference rules except for the modal rule, where it specifies the indices of all the $\lozenge$-formulas that are affected, as well as for the initial rule,
in which case the list contains a single index denoting the complementary literal.

In more technical terms, upon reaching the application of a modal rule in the $OS$ proof evidence, the FPC program generates a proof evidence
in the $\lmfs$ layer which contains a new inference for each $\Diamond$ formula in the $OS$ proof evidence. This is required since
there are no $\Diamond$ inference rules in $OS$. It then populates them with the same
multi-focusing values as well as with the correct futures and presents.

Sec.~\ref{sec:poly} contains more information about the FPC specification.

\subsubsection{Nested Sequents}
\label{sec:ns}

A more challenging example of using our framework is supporting nested sequent proof evidences.
Here we will also demonstrate how layers other than \lmfs\ can be used in order to support proof formats.

Fig.~\ref{fig:nsex} shows a proof of the same theorem we have seen so far, this time by using the $NS$ calculus which was mentioned
in Sec.~\ref{sec:nss}.

\begin{figure}
  \begin{prooftree}
    \AxiomC{$\lozenge(P \wedge \neg Q), \lozenge\neg P, [\neg Q, \neg P,Q]$}
    \AxiomC{$\lozenge(P \wedge \neg Q), \lozenge\neg P, [P, \neg P,Q]$}
    \BinaryInfC{$\lozenge(P \wedge \neg Q), \lozenge\neg P, [P \wedge \neg Q, \neg P,Q]$}
    \UnaryInfC{$\lozenge(P \wedge \neg Q), \lozenge\neg P, [\neg P,Q]$}
    \UnaryInfC{$\lozenge(P \wedge \neg Q), \lozenge\neg P, [Q]$}
    \UnaryInfC{$\lozenge(P \wedge \neg Q), \lozenge\neg P, \square Q$}
    \UnaryInfC{$\lozenge(P \wedge \neg Q), \lozenge\neg P \vee \square Q$}
    \UnaryInfC{$\lozenge(P \wedge \neg Q) \vee (\lozenge\neg P \vee \square Q)$}
  \end{prooftree}
  \caption{Nested sequent proof of axiom $K$.}
  \label{fig:nsex}
\end{figure}

Unlike $OS$ proofs, we can see a closer relationship to the $LS$ calculus, which forms the basis of our framework.
While the $OS$ calculus has one modal rule, the $NS$ calculus rules correspond, more or less to the ones in $LS$.
This allows us to use directly the \lmf\ layer for the certification of $NS$ proofs.

We note though two differences between $NS$ and $LS$. Nested sequents do not use labels and in order to index them, we need more than
just the symbols $\rgt$, $\lft$, $\diaind$ and $\emptyset$. We now index formulas using two
separate indices: The first one is just the location of the sub-formula as before while the second is the index of the nested sequent,
as will be explained next.

We remind the reader that our framework is not based on labels but on correspondences between indices. Therefore, the translation
between $NS$ and \lmf\ will only require a consistent mapping between the indices defined next and those of the \lmf\ layer.

\begin{definition}[Indexing a Nested Sequents]
Indices of nested sequents within a sequent are defined recursively by:
\begin{itemize}
  \item \texttt{zb} is an index (of the top level nested sequent).
  \item if \texttt{ind} is an index of a nested sequent containing the nested sequents $S_1,\ldots,S_n$,
    then \texttt{(chld i ind)} is an index denoting the nested sequent at position $i$.
\end{itemize}
\end{definition}

The index of a sequent in $NS$ will then be composed of a pair containing our regular index and the index which was just defined.

Fig.~\ref{fig:nesex} gives an example of a nested sequent derivation and the indices of sub-formulas.

\begin{figure}
  \begin{prooftree}
    \AxiomC{$(q)^{\texttt{(ns (lind root) zb)}}, ([p])^{\texttt{(ns (rind root) (chld 1 zb))}}$}
    \UnaryInfC{$(q)^{\texttt{(ns (lind root) zb)}}, (\square p)^{\texttt{(ns (rind root) zb)}}$}
    \UnaryInfC{$(q \vee \square p)^{\texttt{(ns root zb)}}$}
  \end{prooftree}
  \caption{An example of a nested sequent derivation and the corresponding indices.}
  \label{fig:nesex}
\end{figure}

In order to certify nested sequent proofs in our framework, we will use, as mentioned above, the \lmf\ layer.
We note that the $NS$ $\Box$ rule creates a new nested sequent while the $NS$ $\Diamond$ rule adds the formula into
an existing nested sequent. This is similar to the correspondence between $\Box$ and $\Diamond$ formulas we have defined
in Def. \ref{df:cor}. Our FPC specification for $NS$ will indeed exploit this similarity and will translate between the indices
of the two systems, $NS$ and \lmf.

Fig.~\ref{fig:nstns} shows the idea behind this translation. In order to keep track of the index translations, we
add to the proof evidence an empty structure which denotes its state. This state is being updated along the certification
process by the FPC specification.

\begin{figure}
\begin{lstlisting}[
  basicstyle=\small\ttfamily
]

nested_to_single
  (nested_cert (nested_state Map) I OI)
  (single_cert I' OI') Map :-
 map_index Map I I',
 map_index Map OI OI'.

single_to_nested Map
  (single_cert I' OI')
  (nested_cert (nested_state Map) I OI) :-
 map_index Map I I',
 map_index Map OI OI'.

orNeg_c Cert Form Cert' :-
 nested_to_single Cert Cert-s Map,
 orNeg_c Cert-s Form Cert-s',
 single_to_nested Map Cert-s' Cert'.
\end{lstlisting}
  \caption{Proof evidence transformation between $NS$ and $\lmf$.}
\label{fig:nstns}
\end{figure}

In general, supporting nested sequent proof evidence for $K$ is straightforward and does not
require any knowledge of \LKF. The only thing required is to be able to translate between the indices.
Our use of the "polymorphic" approach means that understanding of the \LKF\ inference rules is still required.
In Sec.~\ref{sec:conclusion}, we will discuss an alternative approach that eliminates the need to understand the \LKF\ calculus.

\subsection{Examples}
	\label{sec:examples}
In this section, we explain how our program can be executed on different examples. 
%
%
The examples described in the paper and others can be found in the testing section of the \checkers\ proof certifier.
\checkers\ can be obtained on Github\footnote{ \url{https://github.com/proofcert/checkers/tree/dalefest}.}, or on Zenodo\footnote{\url{https://zenodo.org/record/1325924#.W2IWPHVfgWM}}. It depends on the $\lambda$Prolog interpreter Teyjus (\url{http://teyjus.cs.umn.edu/})
and can be executed by running in a bash terminal:
\begin{verbatim}
$ ./prover-teyjus.sh arg
\end{verbatim}
where the argument is the name of the  $\lambda$Prolog module denoting the proof evidence one wishes to check.

We are currently also supporting the ELPI implementation of $\lambda$Prolog\footnote{\url{https://github.com/LPCIC/elpi}}.
For running examples using ELPI, please use:

\begin{verbatim}
$ ./prover-elpi.sh arg
\end{verbatim}

Unfortunately, some bugs in the implementations of $\lambda$Prolog forced us to associate the different examples to specific implementations.
The examples starting with \verb+ex-+ can be executed with Teyjus while the rest are better executed with ELPI.

\section{Discussion and conclusion}
	\label{sec:conclusion}

In this paper, we have presented the implementation of a framework for certifying propositional modal logic proofs. The framework has been developed by following the general principles of the project ProofCert and as a module of the concrete implementation provided by \textsf{Checkers}.
Our approach is based on the use of a layered architecture, which allowed us to design a modular framework capable of supporting many different proof formats.
  While layers provide a high level of flexibility, we observe that the need to translate between proof evidences in order to support
  proof evidence polymorphism gets increasingly more complex, the farther we get from the kernel. In addition, FPC specifications
  are still defined in terms of the \LKF\ kernel, which requires an implementer to be familiar with \LKF.
%
  A different approach would consist in having a distinct (``concrete'') kernel for each layer.
  However this could compromise the trust one can place in such layers and would go against
  the general principle of having kernels based on simple, well-known and low-level calculi.
  In addition, since different proof calculi might use different kernels, the property of proof
  universality would be lost.

  In~\cite{ChiMilRen16}, it has been shown how it is possible to ``host'' the classical calculus $\LKF^a$ on an intuitionistic focused kernel,
  by using a translation between the two logics. Along the lines of what has been proposed there, we are investigating the possibility of representing the different layers as ``virtual'' kernels, built on top of a lower level kernel.
  We briefly illustrate the idea by showing how $\lmf^a$ (a version of $\lmf$ augmented with proper clerks and experts) can be ``hosted'' on $\LKF^a$.
  The calculus $\lmf^a$ is shown in Fig.~\ref{fig:LMFa}. The augmentation leading from $\lmf$ to $\lmf^a$ is similar in spirit to the one going from $\LKF$ to $\LKF^a$ and is obtained by adding control predicates to the base system. We remark that in the case of relational formulas, we do not need to store them with a significant index as we will never focus on them again.
Now we can provide a definition of the clerks and experts of $\LKF^a$ in terms of those given for $\lmf^a$, as shown in Fig.~\ref{fig:lkf-lmf}.
  Like in the approach of Sec.~\ref{sec:cert}, and more generally in any attempt to find a classical first-order proof for a propositional modal formula, an adequate translation is required. The definition of Fig.~\ref{fig:lkf-lmf} goes therefore together with a simple translation from the polarized propositional modal language to the polarized first-order language, which basically maps each classical connective into the corresponding connective
  (by also preserving polarity) and translates $\square$ and $\lozenge$ as in Sec.~\ref{sec:emul-theory}.
  We omit a full type declaration and just remark that in Fig.~\ref{fig:lkf-lmf}, \texttt{C}, \texttt{C'} and \texttt{C''} stand for $\lmf$ evidences, while \texttt{mod} and \texttt{tns} are constructors that, applied to an $\lmf$ evidence, produce an $\LKF$ evidence. Intuitively, we use them to distinguish between two phases along the construction of a proof: a phase dealing with connectives introduced by the translation of $\square$ or $\lozenge$ (denoted by \texttt{tns}) and a ``normal'' one that does not involve the translation of modalities (denoted by \texttt{mod}).
  By using such an inter-definition, an external user interested in certifying her proofs over $\lmf$ can assume that a kernel based on $\lmf^a$ indeed exists and only needs to define $\lmf^a$ clerks and experts.
  In the context of our framework, by relying on the same idea, we could base each kernel on the immediately lower one. This solution would allow for keeping only one trusted kernel - $\LKF^a$ - but would, at the same time, provide virtual kernels which can be used in order to write simpler FPC specifications for the different proof formats.
  As an example, Fig.~\ref{fig:nsvir} shows how an FPC specification can be written for the system $NS$, in a framework that uses virtual kernels, and should be compared with the specification in Fig.~\ref{fig:nstns}.
  As one can see, we can now use directly the clerks and experts of the \lmf\ layer.

\begin{figure}
\begin{lstlisting}[
  basicstyle=\small\ttfamily
]

orNeg_LMFc
 (nested-cert node I O [H|_]))
 (nested-cert (node H)).
\end{lstlisting}
  \caption{Proof evidence transformation between $NS$ and $\lmf$, by using virtual kernels.}
\label{fig:nsvir}
\end{figure}
%



		\begin{figure}[t]\scriptsize
		{\sc Asynchronous introduction rules}
\renewcommand{\Async}[3]{\blue{#1, \,}\ar\vdash#2\mathbin{\Uparrow}   #3}
\newcommand{\Asyncrel}[4]{\blue{#1, \,}\ar\cup\{#2\}\vdash#3\mathbin{\Uparrow}   #4}
\renewcommand{\Sync }[3]{\blue{#1, \,}\ar\vdash#2\mathbin{\Downarrow} #3}
\newcommand{\Syncrel}[4]{\blue{#1, \,}\ar\cup\{#2\}\vdash#3\mathbin{\Downarrow} #4}
\renewcommand{\andClerk}[3]{\blue{{\hbox{andNeg$_c$}}(#1,#2,#3)}}
\renewcommand{\falseClerk}[2]{\blue{\hbox{f$_c$}(#1,#2)}}
\renewcommand{\orClerk}[2]{\blue{{\hbox{orNeg$_c$}}(#1,#2)}}
\renewcommand{\allClerk}[2]{\blue{\hbox{all$_c$}(#1,#2)}}
\renewcommand{\boxClerk}[2]{\blue{\hbox{box$_c$}(#1,#2)}}
\renewcommand{\storeClerk}[4]{\blue{\hbox{\sl store$_c$}(#1,#2,#3,#4)}}
\renewcommand{\trueExpert }[1]{\blue{{\hbox{true$_e$}}(#1)}}
\renewcommand{\andExpert}[3]{\blue{{\hbox{andPos$_e$}}(#1,#2,#3)}}
\renewcommand{\andExpertLJF}[6]{\blue{{\hbox{andPos$_e$}}(#1,#2,#3,#4,#5,#6)}}
\renewcommand{\orExpert  }[3]{\blue{{\hbox{orPos$_e$}}(#1,#2,#3)}}
\renewcommand{\someExpert}[3]{\blue{\hbox{some$_e$}(#1,#2,#3)}}
\renewcommand{\diamondExpert}[3]{\blue{\hbox{dia$_e$}(#1,#2,#3)}}
\renewcommand{\initExpert}[2]{\blue{\hbox{\sl initial$_e$}(#1,#2)}}
\renewcommand{\initrExpert}[1]{\blue{\hbox{\sl initialR$_e$}(#1)}}
\renewcommand{\cutExpert}[4]{\blue{\hbox{\sl cut$_e$}(#1,#2,#3,#4)}}
\renewcommand{\decideExpert}[3]{\blue{\hbox{\sl decide$_e$}(#1,#2,#3)}}
\renewcommand{\releaseExpert}[2]{\blue{\hbox{\sl release$_e$}(#1,#2)}}
\[
\infer{\Async{\Xi}{\Theta}{x:A\wedgen B,\Gamma}}
      {\Async{\Xi'}{\Theta}{x:A,\Gamma} \quad
       \Async{\Xi''}{\Theta}{x:B,\Gamma} \quad
       \andClerk{\Xi}{\Xi'}{\Xi''}}
\]
\[
\infer{\Async{\Xi}{\Theta}{x:A\veen B,\Gamma}}
      {\Async{\Xi'}{\Theta}{ x:A,x:B,\Gamma}\quad\orClerk{\Xi}{\Xi'}}
\qquad
\infer[\dag]{\Async{\Xi}{\Theta}{x:\square B,\Gamma}}
      {\Asyncrel{(\Xi' y)}{\tupp{relind}{R(x,y)}}{\Theta}{y:B,\Gamma}\quad\boxClerk{\Xi}{\Xi'}}
\]
		{\sc Synchronous introduction rules}
		\[
\infer{\Sync{\Xi}{\Theta}{x:B_1\wedgep B_2}}
      {\Sync{\Xi'}{\Theta}{x:B_1}\quad
       \Sync{\Xi''}{\Theta}{x:B_2}\quad
       \andExpert{\Xi}{\Xi'}{\Xi''}}
\]
\[
\infer{\Sync{\Xi}{\Theta}{x:B_1\veep B_2}}{\Sync{\Xi'}{\Theta}{x:B_i}\qquad
       \orExpert{\Xi}{\Xi'}{i}}
\qquad\qquad
\infer{\Syncrel{\Xi}{\tupp{relind}{R(x,y)}}{\Theta}{x:\lozenge B}}{\Syncrel{\Xi'}{\tupp{relind}{R(x,y)}}{\Theta}{y:B}\quad
                  \diamondExpert{\Xi}{y}{\Xi'}}
\]
		{\sc Identity rules}
\[
\infer[init]{\Sync{\Xi}{\Theta}{x:B}}
            {\tupp{l}{x:\neg B}\in\Theta\quad\initExpert{\Xi}{l}}
\]
		{\sc Structural rules}
\[
\infer[\kern -1pt release]{\Sync{\Xi}{\Theta}{x:B}}
               {\Async{\Xi'}{\Theta}{x:B}\quad\releaseExpert{\Xi}{\Xi'}}
\qquad
\infer[store]{\Async{\Xi}{\Theta}{x:B,\Gamma}}
             {\Async{\Xi'}{\Theta, \tupp{l}{x:B}}{\Gamma} \quad
              \storeClerk{\Xi}{x:B}{l}{\Xi'}}
\]
\[
\infer[\kern -1pt decide]{\Async{\Xi}{\Theta}{\cdot}}
              {\Sync{\Xi'}{\Theta}{x:B}\quad
               \tupp{l}{x:B}\in\Theta\quad
               \decideExpert{\Xi}{l}{\Xi'}}
\]
	In $decide$, $B$ is positive; in $release$, $B$ is negative; in $store$, $B$ is a positive formula or a negative literal; in $init$, $B$ is a positive literal. The proviso $\dag$ specifies that $y$ is different from $x$ and does not occur in $\Theta$ nor in $\ar$.
		\caption{$\labkF^a$: a focused labeled proof system for the modal logic $\logick$.}
		\label{fig:LMFa}
		\end{figure}

\begin{figure}
\begin{lstlisting}[
  basicstyle=\small\ttfamily
]

andNeg_LKFc (mod C) (mod C') (mod C'') :- andNeg_LMFc C C' C''.
orNeg_LKFc (tns C) (tns C).
orNeg_LKFc (mod C) (mod C') :- orNeg_LMFc C C'.
all_LKFc (mod C) (X\ tns (C' X)) :- box_LMFc C C'.

andPos_LKFe (tns C) (tns C) (mod C).
andPos_LKFe (mod C) (mod C') (mod C'') :- andPos_LMFe C C' C''.
orPos_LKFe (mod C) (mod C') LeftRight :- orPos_LMFe C C' LeftRight.
some_LKFe (mod C) Term (tns C') :- dia_LMFe C Term C'.

initial_LKFe (mod C) Index :- initial_LMFe C Index.
initial_LKFe (tns C) relind.

release_LKFe (mod C) (mod C') :- release_LMFe C C'.
store_LKFc (tns C) relind (mod C).
store_LKFc (mod C) Index (mod C') :- store_LMFc C Index C'.
decide_LKFe (mod C) Index (mod C') :- decide_LMFe C Index C'.
\end{lstlisting}
  \caption{The definition of
  $\LKF^a$ clerks and experts based on those given for
  $\lmf^a$.}
\label{fig:lkf-lmf}
\end{figure}

Besides further investigation in this direction, there are several ways in which this work can be extended.
The design of the parametric devices of the framework has been driven by the ambition of being as comprehensive as possible in terms of formalisms captured. The modularity and parameterizability of the whole approach should make it possible, in fact, to consider other related approaches to modal proof theory, like hypersequent calculi~\cite{Avr94}, e.g.,~by using a \emph{present} parameter that is a multiset, representing external structural rules as operations on such a
present, and viewing modal communication rules as a combination of
relational and modal rules.
The focused nature of the approach should also allow for certifying proofs coming from focused proof systems for modal logics, like the ones in~\cite{LellmannP15,Cha16}, possibly by using a different polarization of formulas.

Orthogonally, we also aim at extending the approach to variants of the logic K.
By applying the results in~\cite{MarMilVol16}, the extension to the logics characterized by the so-called geometric frames seems to be not too complex, although it might require some modifications to the current implementation (e.g., in the case of a logic defined, amongst the others, by the axiom of seriality $D$, a more complex notion of modal correspondence between $\lozenge$-formulas and $\square$-formulas, or labels, would be required).

We also notice that while this work was inspired by certification consisting in a strict emulation of original proofs, it is sometimes the case that only partial information about the proof to be checked is provided. We plan to complement the current implementation with a ``relaxed'' version of the FPCs, such that it can also deal with incomplete proof evidences, similarly to what has been done in~\cite{LibalV16} in order to check, e.g.,~free-variable tableau~\cite{Beckert97a} proofs.

The kind of investigation done in this work also suggests us new directions to explore, more generally, in the context of the \checkers\ project. For instance, it seems interesting to consider the application of \checkers\ to objects composed from proofs coming from different proof calculi. One example of such objects
are coalesced proofs, described in \cite{tla14}. In this work, a proof evidence is created by using modal theorem provers alongside first-order ones.
A universal proof certifier such as \checkers, which is based on \LKF, can attempt to use the different components in order to find
a composed formal proof of the original theorem.
One can also compare this goal to the one achieved by different ``hammers''
\cite{hammers13}, where specialized theorem provers are indirectly used
in order to find a formal proof in a different calculus. The success and popularity of the different hammers makes a case
in favor of further extensions of \checkers. An example of such an extension, beyond first-order and propositional modal proofs, is
the support of proofs based on theories, such as arithmetic.

\smallskip
\noindent{\bf Acknowledgment.} The work presented in this paper was partially funded by the ERC Advanced Grant Proof\kern 0.6pt Cert. Part of the work was carried out while the authors were at INRIA Saclay.

%

\label{sect:bib}
\bibliographystyle{plain}

\end{document}